\documentclass[journal]{IEEEtran}
\usepackage{epsfig,graphics,graphicx,amssymb,amstext,amsmath,algorithm,
algorithmic,wrapfig,multirow}
\usepackage{cite}
\usepackage{url}
\usepackage{times}
\usepackage{float}
\usepackage{placeins}
\usepackage{textcomp}
\usepackage[font=small,bf]{caption}
\usepackage{flushend}
\usepackage{xcolor}
\usepackage{subcaption}
\usepackage{algorithm}
\usepackage{authblk}
\usepackage{bbm}
\usepackage{dsfont}
\usepackage{balance}
\AtEndDocument{\par\leavevmode}

\usepackage{tabularx}
\usepackage{perpage} 

\MakePerPage{footnote} 

\IEEEoverridecommandlockouts

\renewcommand{\footnoterule}{%
  \kern -3pt
  \hrule width \columnwidth height 0.5pt
  \kern 3pt
}
\begin{document}

\title{SVC-based Multi-user Streamloading for Wireless Networks}
\author[1]{S. Amir Hosseini}
\author[2]{Zheng Lu}
\author[2]{Gustavo de Veciana}
\author[1]{Shivendra S. Panwar}

\affil[1]{Department of Electrical and Computer Engineering, NYU Tandon School of Engineering}
\affil[2]{Department of Electrical and Computer Engineering, University of Texas at Austin}

\maketitle

\begin{abstract}
In this paper, we present an approach for joint rate allocation and quality 
selection for a novel video streaming scheme called 
\textit{streamloading}. Streamloading is a recently developed method for delivering high quality video without 
violating copyright enforced restrictions on content access for video streaming. In 
regular streaming services, content providers restrict the amount of viewable video that users can download prior to playback. This approach can cause inferior user experience due to bandwidth variations, 
especially in mobile networks with varying capacity. In streamloading, the video is 
encoded using Scalable Video Coding, and users are allowed to pre-fetch enhancement layers and store them on the  
device, while base layers are streamed in a near real-time fashion 
ensuring that buffering constraints on viewable content are met.

We begin by formulating the offline problem of jointly optimizing
rate allocation and quality selection for streamloading in a wireless 
network. This motivates our proposed online algorithms for joint scheduling at the base station and segment quality selection at receivers. The results indicate 
that streamloading outperforms state-of-the-art streaming schemes  in terms 
of the number of additional streams we can admit for a given video quality. Furthermore, the quality adaptation mechanism of our proposed 
algorithm achieves a higher performance than baseline algorithms with no (or limited) video-centric optimization of the base station's allocation of resources, e.g., proportional fairness.
\end{abstract}


\section{Introduction}\label{sec:intro}
Mobile video streaming services continue to gain  
popularity among cellular data users. 
Currently, video traffic has the largest share of 
cellular data (55\% at the end of 2014), and this trend is predicted to continue 
growing \cite{cisco2015}. In order 
to efficiently meet this demand with the limited bandwidth resources
of wireless networks, the use of video quality adaptation
has gained enormous interest in industry.

Adaptive video transmission over HTTP has been standardized 
under the commercial name DASH \cite{stockhammer2011dynamic}, where 
the video is divided into segments, and multiple 
versions of each segment are encoded at
different bit rates. When a segment is to be downloaded
for viewing, a decision is made based on the conditions in the 
network or on the state of the receiver download buffer,
regarding which segment representation to retrieve. Other video delivery systems use 
scalable video coding
(SVC), an extension of the H.264/AVC standard. 
In SVC, rather than encoding each segment into multiple versions with different bit
rates, the video segments are encoded into several streams called layers. The base
layer may be encoded as a low-quality video, while additional
enhancement layers provide incremental improvements in quality.
This delivery scheme offers additional 
flexibility over DASH, and opens up new options to improve 
video delivery and network efficiency.

In the context of copyrighted video streaming, content 
owners tend to provide {\em conditional access} to users in order to 
tightly control the content being watched, prevent illegal distribution 
of content, implement smart content pricing mechanisms, etc. One of the
most 
widely used conditional access schemes is to limit the amount of viewable
video that can be pre-fetched and stored on the end user device ahead 
of the playback. This limit is specified in the license agreements 
between content owner and content distributor, and varies from tens of 
seconds to a few minutes.

Based on this, we distinguish 
between two service models for video delivery in wireless networks. 
We refer to \textit{streaming} as a service model where only 
a limited number of video segments can be delivered to the user ahead of playback. Streaming services are usually inexpensive 
and content providers tend to monetize the service by injecting 
advertisements during playback or through subscription models. Due to the limited buffering, streaming
may suffer from degraded quality of service under varying channel conditions.
Apart from this, streaming license agreements prohibit making copies or distribution of video content \cite{netflix_st}.
We refer to \textit{downloading} as a service model in 
which the amount of buffered video is not limited.
Unlike streaming, a persistent Internet connection is not necessary 
during playback and users can watch the downloaded video at any, possibly constrained, future 
time. Since users are allowed to store copies of purchased content on their devices  \cite{amazon_dl}, this service model is subject to additional licensing restrictions 
such as the duplication license, and is therefore offered at 
significantly higher prices, typically two orders of magnitude higher than streaming. It should be noted that even if the downloaded content is
encrypted, it falls under this category. Despite the higher
price, downloading offers higher video quality to the user and also improves the efficiency of the data transmission, since there are no buffering constraints.

Recently, a hybrid service model for video delivery  was proposed called \textit{streamloading} \cite{rath_streamloading_2013}. In 
this scheme, the video is encoded into several layers using
SVC, and the base 
layer is streamed in real-time with limited buffering 
at the end user device; while enhancement 
layers may be downloaded ahead of time without buffering 
restrictions. Using this approach, the video quality is 
improved because the receiver can take advantage of available excess bandwidth to download enhancement layers associated with future segments, 
thereby smoothing the effect of variations in link capacity. 
Consequently, users may enjoy video quality similar to that of a downloading 
service, while still being classified 
as a streaming service from the content provider’s point of view \cite{law}. The latter stems from the fact that 
enhancement layers cannot be decoded and are therefore of no value, unless the 
respective base layer is available \cite{schwarz2007overview} \footnote{Downloading encrypted video, which is not viewable without, for instance, an encrypted stream of keys, is still legally classified as downloading, and cannot be classified as a streaming service \cite{law}.}. 

The main contributions of this paper are as follows. We first formulate an optimization framework to study the joint base station rate allocations and segment quality selection problem for a multi-user setting. Leveraging previous work on optimizing DASH-based algorithms, we propose the first comprehensive solution for streamloading. Our results show that streamloading provides significant benefits, e.g., high video quality and low re-buffering time, suggesting that this service model has the potential of providing high Quality of Experience (QoE) while meeting the legal requirements of a streaming service.


The rest of the paper is organized as follows. Section 
\ref{sec:rel_work} contains a summary of existing research on 
optimal adaptive video delivery in multi-user networks. In 
Section 
\ref{sec:access}, we provide a detailed description  of the three
service models mentioned above. The system characteristics, as well as the 
video quality model, are discussed in Section \ref{sec:form} along 
with an offline optimization formulation for multi-user 
streamloading. Section \ref{sec:sol} contains the proposed online
RAte allocation and QUality sELection algorithm (RAQUEL), followed by a discussion of practical implications of our proposed scheme on real networks in Section \ref{sec:prac}. A thorough simulation analysis is presented in Section 
\ref{sec:sim}. Finally, Section \ref{sec:conc} concludes the paper
and briefly discusses potential future research opportunities.

\section{Related Work} \label{sec:rel_work}

A great deal of research has focused on optimal 
resource allocation and quality adaptation for video delivery in 
wireless networks, see, e.g., \cite{zhou2014control,ma2011http,jiang2012improving,
joseph2013nova} and references therein. A large portion of this research deals with  the algorithms and performance of DASH-based video delivery. Bandwidth management for live 
streaming HTTP-based applications is studied in \cite{ma2011http}. Several commercial adaptive video 
streaming services are compared in \cite{jiang2012improving} in terms of bandwidth 
utilization, fairness, and bit rate stability.
The authors conclude that all current services fail to satisfy one or more of 
these requirements, and claim that a randomized 
scheduling and state dependent rate adaptation approach outperforms currently used services. 
In \cite{zhou2014control}, the authors consider the problem of optimal 
rate adaptation of DASH video transmission from multiple content 
distribution networks. In order to keep the bit rate stable, they propose to perform block level rate 
allocation, where multiple segments are grouped together and are transmitted at the 
same bit rate. The work in \cite{joseph2013nova} 
formulates the problem of optimal delivery of DASH-based video 
to wireless users as a dynamic network utility (video quality) maximization problem with 
re-buffering and delivery cost constraints. Based on this formulation, they 
develop an online algorithm called NOVA, which they prove to be asymptotically optimal in stationary regimes. 

As scalable video gains acceptance, particularly after its inclusion in the new H.265(HEVC) \cite{vidyo_vidyo_2012} and VP9 \cite{vidyo_2013} codecs, more research 
work is being dedicated to optimizing SVC delivery, especially in 
wireless networks. Several works have investigated the benefits 
of using SVC over AVC in terms of caching efficiency and adaptation performance \cite{sanchez2012efficient}, as well as 
reduced congestion, especially at the video server end 
\cite{sanchez2011idash}. The problem 
of optimal rate allocation at the base station for SVC video streaming
in a wireless multi-user scenario is investigated in 
\cite{sieber2013implementation,zhaoqoe,talebi2010utility,
zhang2010cross,kim2005optimal}. In \cite{zhang2010cross}, the authors model the quality-rate trade-offs for SVC using a piece-wise linear function, and derive a rate allocation scheme for fading wireless channels. In a similar study, a 
multi-modal sigmoid approximation is used to model the 
quality-rate trade-off where a utility-proportional optimization flow 
control method is used to achieve convexity 
\cite{talebi2010utility}. The mapping of SVC 
layers into DASH representations is studied in \cite{zhaoqoe}. In 
the same work, an optimal scheme for rate allocation and quality
stabilization is proposed for wireless Orthogonal Frequency Division Multiple Access (OFDMA) systems using the 
Lagrangian dual decomposition method. 
A heuristic segment selection approach is used in 
\cite{sieber2013implementation} to dynamically select quality 
layers solely based on buffer level, without considering the 
link bandwidth. In an attempt to decrease the re-buffering time
of SVC delivery, \cite{schierl2011priority} presents a priority 
scheme, in which the base layer segments
are pre-fetched, and enhancement
layers are subsequently downloaded in order to increase video quality. To our knowledge, no comprehensive solution for joint resource allocation and quality selection has been proposed for a multi-user setting, and in particular for streamloading. This is the gap we attempt to fill in this paper.  
\begin{figure*}[t]
\centering
		\captionsetup[subfigure]{twoside,margin={0.5cm,0.5cm}}
		\begin{subfigure}[h]{0.25\textwidth}
             \includegraphics[height=1.5in]{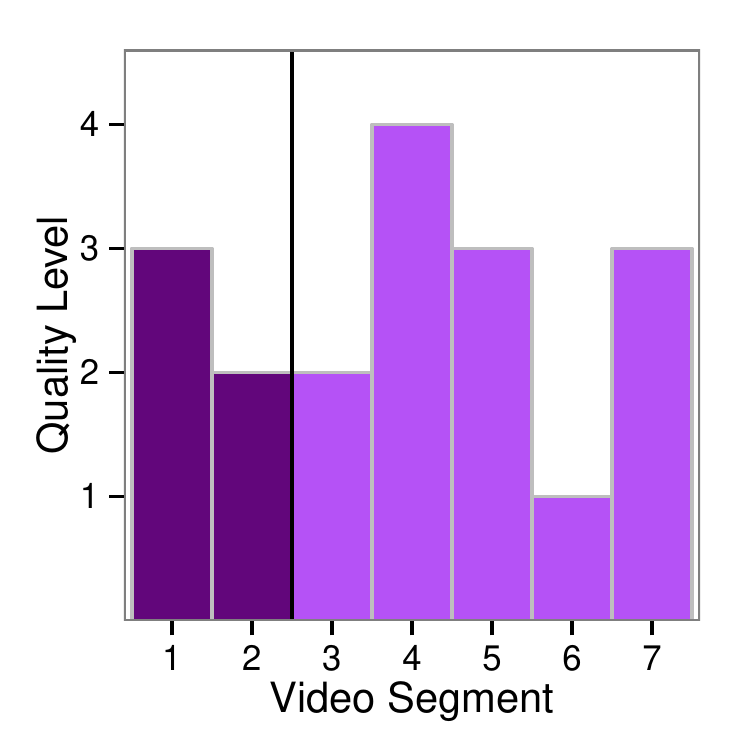}
             \caption{Downloading}\label{fig:access_dl}         
           \end{subfigure}\qquad
           \begin{subfigure}[h]{0.25\textwidth}
             \includegraphics[height=1.5in]{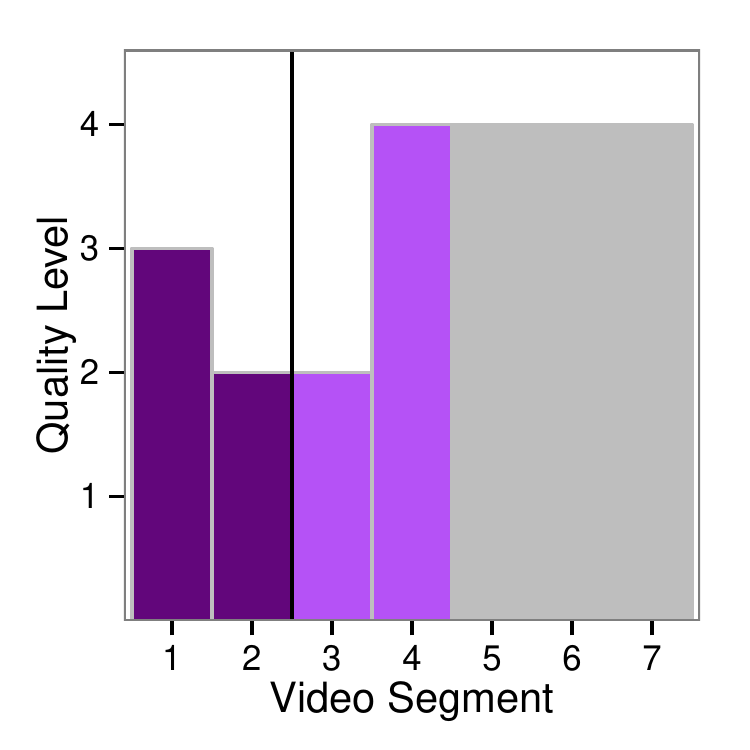}
             \caption{Streaming}\label{fig:access_st}
           \end{subfigure}\qquad       
		\begin{subfigure}[h]{0.25\textwidth}
             \includegraphics[height=1.5in]{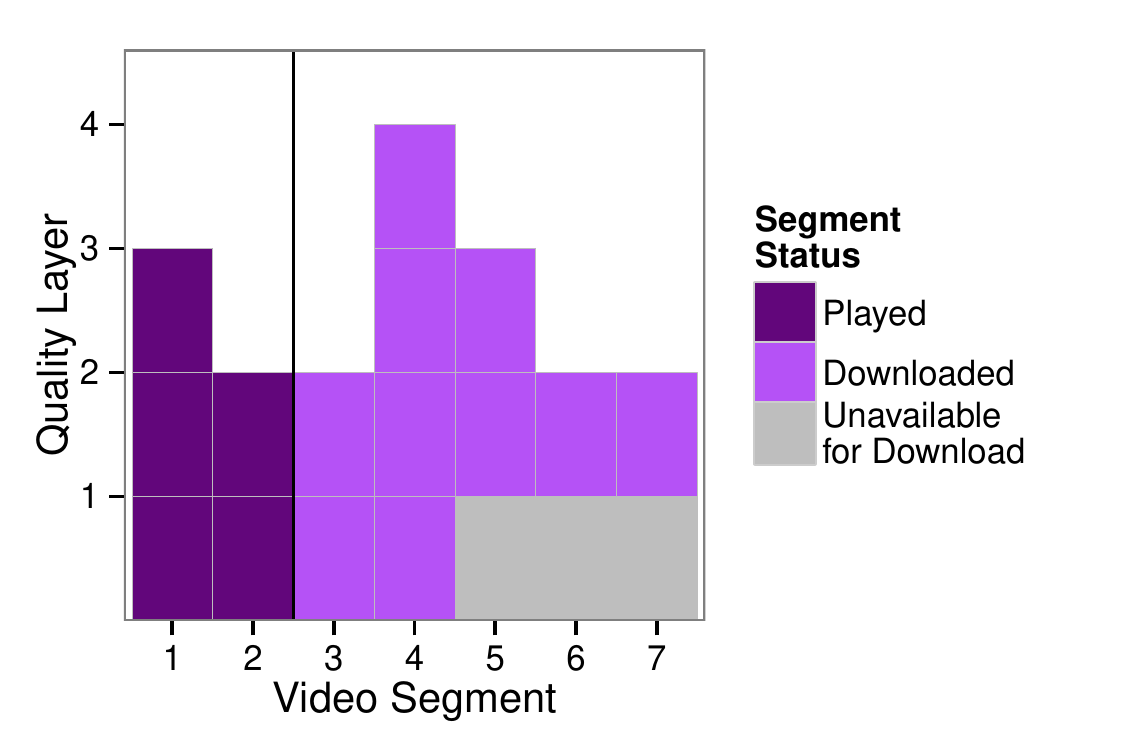}
             \caption{Streamloading}\label{fig:access_sl}         
           \end{subfigure}
\caption{A sample snapshot of the end user buffer for 
downloading, streaming, and streamloading access 
models. The streaming and streamloading service models are 
shown with a 2 segment buffer limit.}
\label{fig:access}         
\end{figure*}

\section{Service models} \label{sec:access}
In this section, we define three different access schemes for 
wireless video delivery in terms of their associated restrictions on the amount
of pre-fetched video content. A graphical illustration of these service models is shown in Figure \ref{fig:access}.

\textbf{Downloading}: This model refers to the case where there is
no restriction on the number of video segments that can be
pre-fetched in advance. In this work, we 
consider a DASH-based delivery for this access scheme, in which 
the video segment are encoded at multiple quality levels. Upon delivery of each segment, the next segment is requested by 
the user. In Figure \ref{fig:access_dl}, an example is
shown for downloading a video sequence with four quality 
representations. It is important to note that here we are talking about short-term pre-fetching. It should 
not be confused with downloading videos overnight and
viewing them later. Indeed, in such a scenario, the video could be delivered at the highest available quality.

\textbf{Streaming}: This service model differs from downloading in that no more than a certain number of segments
can be stored on the user device ahead of playback.
We refer to this as a \textit{buffer limit} from now on. 
Whenever this limit is reached, no further video segments can be received 
until some of the stored video is consumed. An example of 
streaming is shown in Figure \ref{fig:access_st} for a video sequence
with four different representations and with a buffer limit 
of two segments. 

\textbf{Streamloading}: As described in Section \ref{sec:intro}, this 
service model uses SVC to encode each segment into multiple
quality layers, and each layer can be transmitted independently. 
According to this scheme, base layer segments can only be pre-fetched
up to a pre-defined buffer limit. However, the enhancement layer segments
are not subject to such limitations. This scheme is 
illustrated in Figure \ref{fig:access_sl}.

\section{Problem Formulation}\label{sec:form}
In this section, we provide a detailed description of the multi-user streamloading problem and consider an offline optimization framework to explore what an ideal clairvoyant algorithm would aim to do. Table \ref{tab:notation} shows the definition of all the notation used throughout the formulation.

\begin{table} 
\caption {Notation table} \label{tab:title}
\begin{center}
\begin{tabular}{|p{1cm}|p{5cm}|}
\hline
Variable & Definition\\
\hline
$N$ & number of users\\
\hline
$K$ & estimated number of total time slots \\ 
\hline
$S$ & total number of video segments \\ 
\hline
$L$ & total number of enhancement layers for each segment \\ 
\hline
$\tau_{slot}$ & length of time slot\\
\hline
$\tau_{seg}$ & length of video segment\\
\hline
$\tau_{lim}$ & length of buffer limit \\
\hline
$r_{i,k}$ & rate allocated to user $i$ in time slot $k$ \\
\hline
$r_{i,k}^b$ & rate allocated to user $i$ in slot $k$ to download base layer \\ 
\hline
$r_{i,k}^e$ & rate allocated to user $i$ in slot $k$ to download enhancement layer \\ 
\hline
$x_{i,s,k}^b$ & fraction of $r_{i,k}^b$ to download segment $s$\\ 
\hline
$x_{i,s,k}^e$ & fraction of $r_{i,k}^e$ to download segment $s$\\ 
\hline
$q_{i,s}^l$ & quality corresponding to delivering the first $l$ enhancement layers for segment $s$ to user $i$ \\
\hline
$\overline{\beta}$ & maximum value allowed for $\beta_i~\forall i\in\mathcal{N}$\\
\hline
$\beta_i$ & estimated fraction of average re-buffering time to total download time for user $i$ \\ 
\hline
$d_i$ & estimated time to download the entire video for user $i$ \\
\hline
$c_k(\cdot)$ & convex rate region in time slot $k$\\
\hline
$f_{i,s}(\cdot)$ & quality-rate trade-off function for user $i$ and segment $s$\\
\hline
$m_i^S(\cdot)$ & average quality of entire video for user $i$\\
\hline
$v_i^S(\cdot)$ & quality variation of entire video for user $i$\\ 
\hline
$\eta$ & weight of variability in objective function \\ 
\hline
$\gamma_{i,k}$ & number of time slots up to slot $k$ that user $i$ spent re-buffering \\ 
\hline
$S_{i,k}^b$ & total number of base layers fully delivered to user $i$ by slot $k$ \\ 
\hline
$S_{i,k}^e$ & total number of enhancement layers fully delivered to user $i$ by slot $k$ \\ 
\hline
$\mathcal{A}_{i,s,k}^b$ & set of rates allocated to user $i$ to fully download base layers by slot $k$ \\ 
\hline
$\mathcal{A}_{i,s,k}^e$ & set of rates allocated to user $i$ to fully download enhancement layers by slot $k$\\
\hline
\end{tabular}\label{tab:notation}
\end{center}
\end{table}

\subsection{System Model}\label{sec:env_set}
For simplicity, we consider a wireless network consisting of a base station and a 
set of active mobile users $\mathcal{N}$, where $|\mathcal{N}|=N$. 
Time is assumed to be slotted with a slot duration of 
$\tau_{slot}$.
The users are viewing videos, each of which 
is divided into a sequence of segments of 
equal duration $\tau_{seg}$. Each segment is encoded into one
base layer and $L$ enhancement layers. 
Our goal is to develop a scheme for joint base station scheduling and segment quality
selection for ``optimal" streamloading subject to a receiver base layer buffer limit of 
$\tau_{lim}$ seconds.

We let $K$ denote an estimate for the number of time slots required to deliver the entire video to all users. In each time slot $k$, the base station allocates the rate $\mathbf{r}_k$ to all users, where $\mathbf{r}_k = 
(r_{i,k})_{i\in\mathcal{N}} \in \mathbb{R}_{+}^N$. 
The resource allocation is subject to time varying constraints
determined by the link quality and achievable data rate. Therefore, the data rate that the base station can 
allocate to each user in time slot $k$ is restricted to a possibly time varying convex rate region 
defined by $c_k (\mathbf{r}_k) \leq 0$, where $c_k$ is assumed to be a real valued 
(continuous) convex function reflecting constraints on network 
resource allocation in slot $k$. We refer to this as the allocation 
constraint in slot $k$. This model encompasses a wide
range of wireless systems \cite{joseph2013nova}.

We denote by $q_{i,s}$, the perceived video quality achieved by user $i$ for segment $s$.
The quality of a segment in a scalable coded video 
increases with the number of successfully downloaded enhancement layers. Therefore, 
$q_{i,s}$ can take any value from the discrete set $\mathcal{Q}_{i,s}=\{q^0_{i,s},
\cdots,q^L_{i,s}\}$, where $q^l_{i,s}$ is the perceived quality 
obtained from
delivering the first $l$ enhancement layers for segment $s$.
The more enhancement layers the user downloads for a segment, 
the higher the required video data rate will be. We denote the average data
rate associated with segment $s$ downloaded by user $i$ with quality 
level $q_{i,s}$ as $f_{i,s}(q_{i,s})$ in bits per second.
Further, the segment data rate for a 
particular representation $f_{i,s}(.)$ is a convex function 
of the video quality, i.e., the relationship between quality and required video data is concave \cite{joseph2012jointly}. We refer to this convex function
as the \textit{quality-rate} trade-off. It should be noted that higher enhancement layers cannot be decoded if lower enhancement layers are not delivered. 

\subsection{Mathematical Formulation}\label{sec:math_form}

The main objective of our streamloading problem is to maximize the overall video quality experienced by the users. We consider the average video 
quality along with the temporal quality variations as the key factors affecting the overall video quality. Therefore, 
the video quality of user $i$ is calculated as $m_i^S(q_i)- \eta\text{v}^S_i(q_i)$
where, $m_i^S(q_i) = \frac{\sum_{s=1}^S q_{i,s}}{S}$ represents 
the average video quality for user $i$ after receiving the entire $S$ segment long video, and $\text{v}_i^S(q_i) = \sqrt{\frac{\sum_{s=1}^S (q_{i,s} - m_i^S(q_i))^2}{S}}$ is the quality variation for the same
sequence of segments. The weight of quality variability on 
the overall video quality is determined by the 
constant $\eta$. A small value, indicates 
that the quality depends less on the temporal variability and more 
on the average quality, and vice versa. The request of which quality representation to request next is sequentially made by the user.
For each segment $s$, we define the quality vector $\mathbf{q}_i = (q_{i,s})_{s\in\mathcal{S}}$ as the sequence of requested quality levels by user $i$, where $\mathcal{S} = \{1,\cdots,S\}$. Hence, $\mathbf{q}_i$ is the decision variable of user $i$.

The resource allocation is performed at the base station and is subject to the channel capacity constraint as discussed in Section \ref{sec:env_set}. 
Since in streamloading, base and enhancement layers can be scheduled and delivered separately, the base station scheduler has the capability to decide how to allocate rate not just among users, but also among base and enhancement layer segments of each user. Therefore, in order to differentiate the rates allocated to different layers, we denote the rate at slot $k$ as $\mathbf{r}^b_k = (r^b_{i,k})_{i\in\mathcal{N}}$
 and $\mathbf{r}^e_k = (r^e_{i,k})_{i\in\mathcal{N}}$ which correspond to the rate dedicated to base and enhancement layer segments, respectively. Hence, $\mathbf{r}^b_k$ and $\mathbf{r}^e_k$ are the decision variables of the base station at every time slot $k$. 

\begin{figure*}[t]
\centering
		\captionsetup[subfigure]{twoside,margin={1cm,1cm}}
		\begin{subfigure}[h]{0.29\textwidth}
             \includegraphics[height=2.05in]{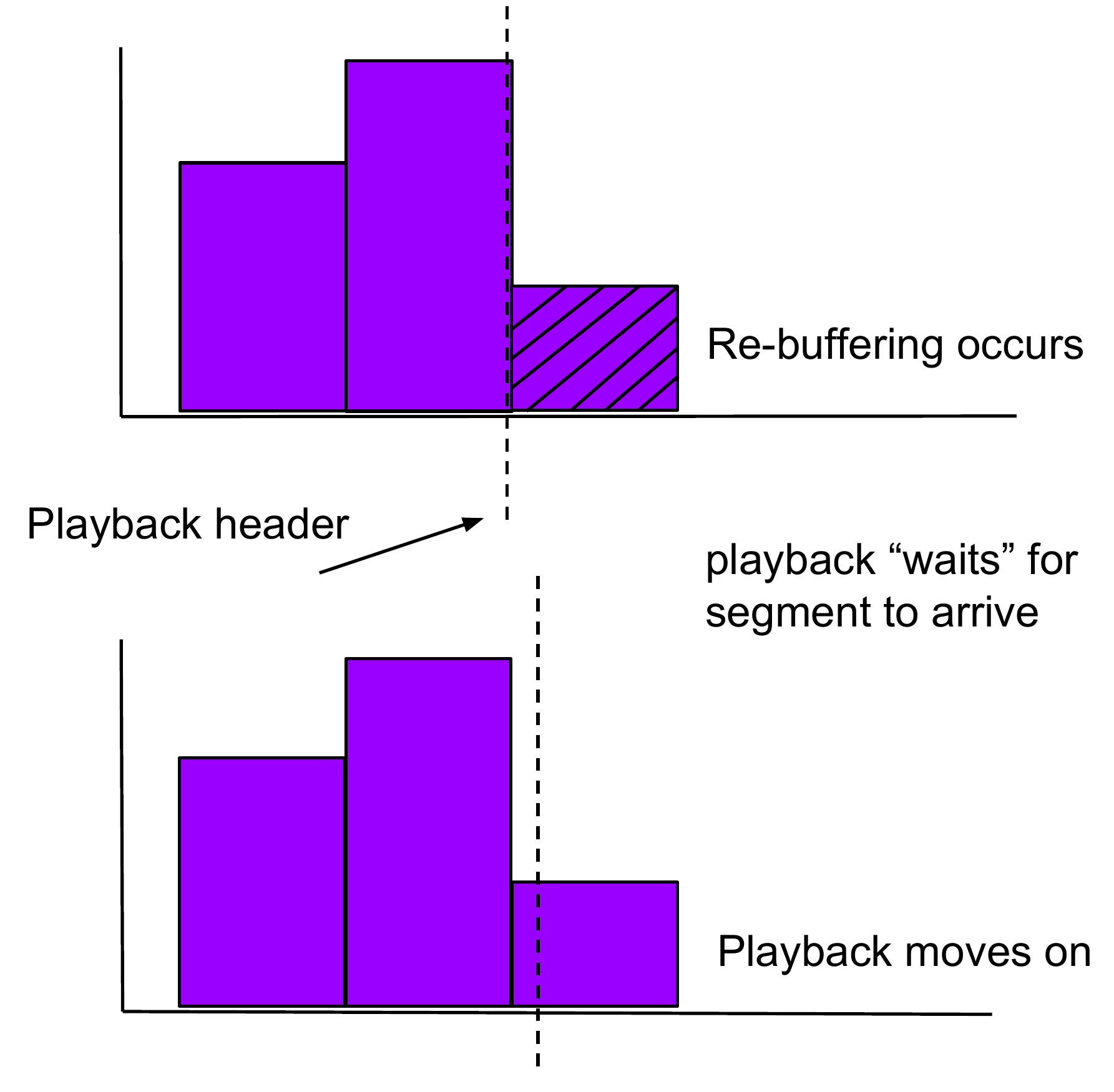}
             \caption{Re-buffering for DASH streaming}\label{fig:rebuf_dash}         
           \end{subfigure}\qquad
           \begin{subfigure}[h]{0.29\textwidth}
             \includegraphics[height=1.96in]{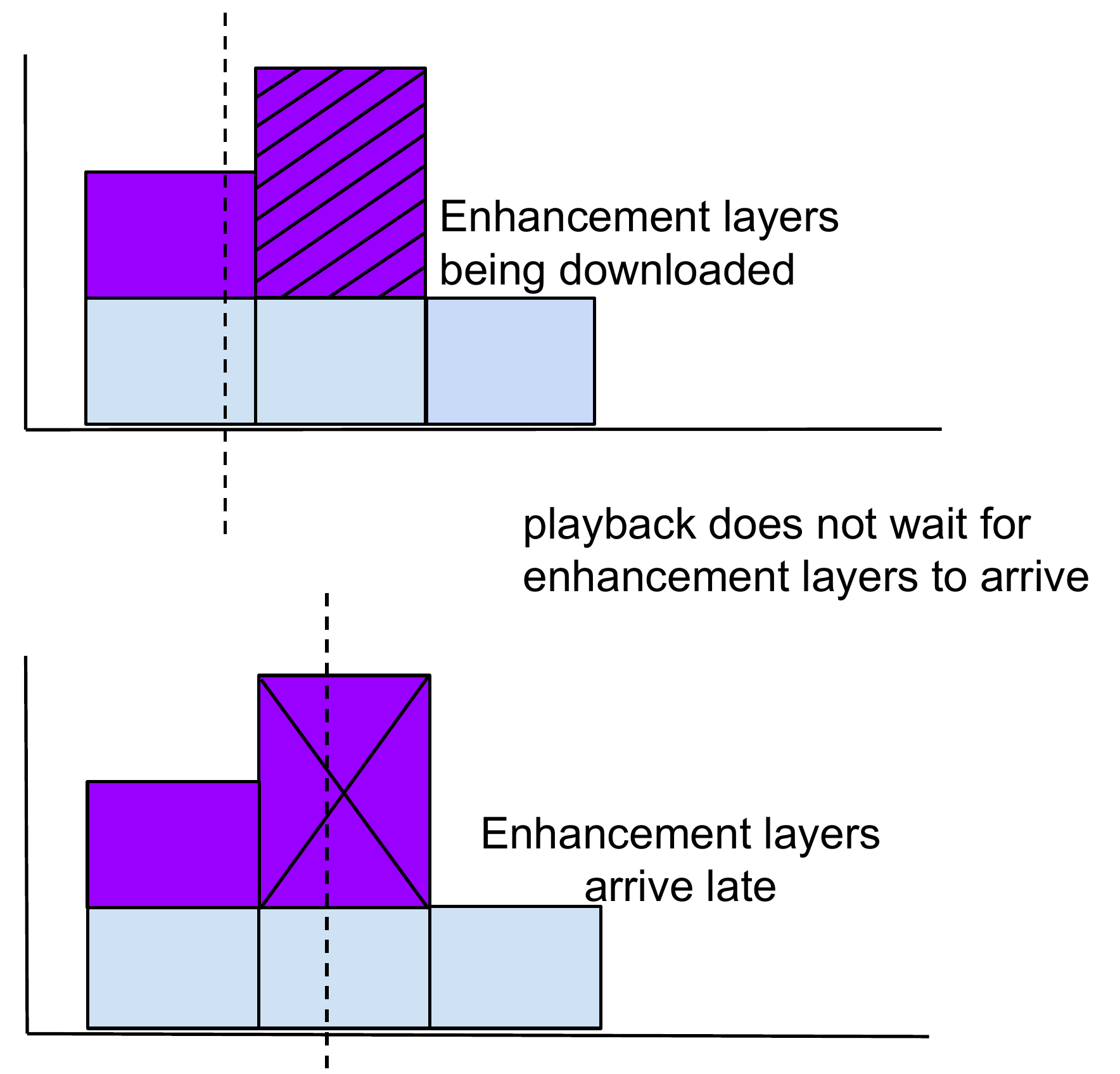}
             \caption{Segment loss in streamloading }\label{fig:sl_late}
           \end{subfigure}\qquad       
		\begin{subfigure}[h]{0.31\textwidth}
             \includegraphics[height=2.09in]{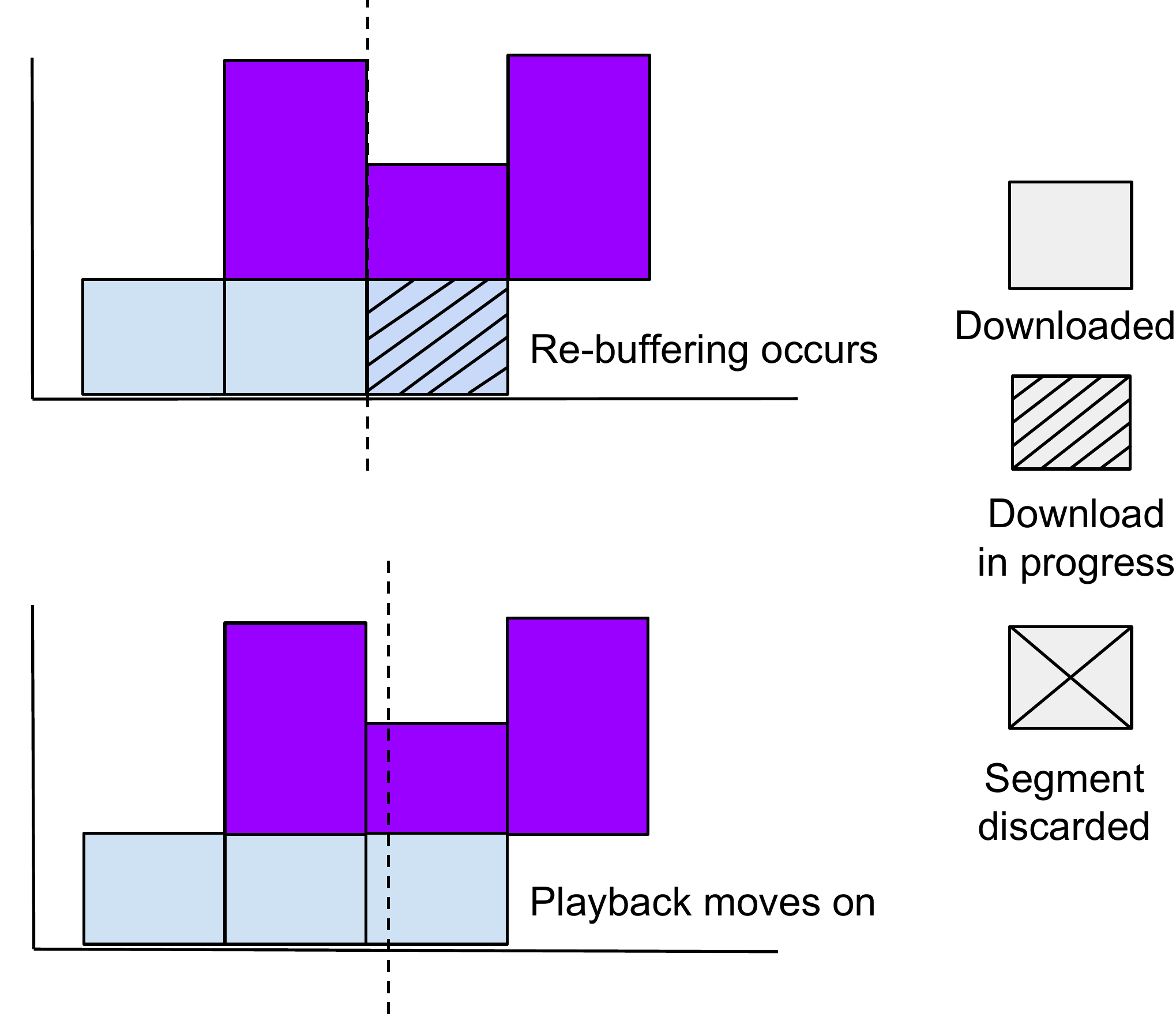}
             \caption{Re-buffering in streamloading}\label{fig:rebuf_sl}         
           \end{subfigure}
\caption{This figure illustrates the different behavior between
streamloading and conventional DASH based streaming in terms of re-buffering and segment late arrival. In each pair, the upper figure shows the buffer at a given time instant and the lower one shows the same buffer at a later point in time. The dashed vertical line indicates the location of the playback. It can be seen that in DASH streaming, the playback stops and re-buffers whenever the segment to be played back has not fully arrived. In streamloading, depending on whether the segment that has not been fully delivered for playback is a base, or enhancement layer, re-buffering, or segment loss, respectively, will occur.}
\label{fig:rebuf}         
\end{figure*}

Ideally, all users prefer to receive the full quality for all segments. However, because of the channel capacity constraint, increasing the load on the network causes delay in delivering video segments. This delay can result in {\em re-buffering}, which manifests itself to the user as a frozen video frame and causes major degradation to the QoE of video streaming. Therefore, quality selection mechanisms should limit re-buffering. In streamloading, since base and enhancement layer delivery is decoupled, the analysis of re-buffering is different than in single layered DASH as shown in Figure \ref{fig:rebuf}.
In DASH video delivery, a segment 
is played back only if it has been fully delivered to the receiver. 
If the playback time reaches a segment which has not been completely 
delivered, re-buffering occurs and the playback stops until the delivery is complete as illustrated in Figure \ref{fig:rebuf_dash}. However, since in SVC, base layers can be decoded and played 
back with or without enhancement layers, the re-buffering time is 
solely determined by the time that the base layer buffer is empty. An example for re-buffering in streamloading is shown in Figure \ref{fig:rebuf_sl}. In order to limit the average re-buffering time of a client, we set an upper bound on the 
fraction of playback time that users experience re-buffering.

If we let $d_i$ denote an estimate for the total time required to download $S$ base layer segments for user $i$, we have:

\begin{equation}\label{eq:dl_time}
d_i=\frac{\tau_{seg}\sum_{s=1}^S f_{i,s}(q^0_{i,s})}{\frac{1}{K}\sum_{k=1}^{K}r^b_{i,k}},
\end{equation}
which is simply the total delivered base layer data over the average allocated rate to that user.
For very large $S$ in a stationary regime, 
the denominator in (\ref{eq:dl_time}) gives an estimate for the 
average base layer download rate of user $i$ over a long period. 
Consequently, one can estimate the fraction of time that user $i$ is 
re-buffering, which we denote as $\beta_i$, as 
$\frac{d_i}{\tau_{seg}S}-1$, 
We can rewrite $\beta_i$ as follows:

\begin{equation}\label{eq:rebuf}
\beta_{i}(\mathbf{q}_i,(r^b_{i,k})_{k\in\mathcal{K}}) := \frac{\sum_{s=1}^S f_{i,s}(q^0_{i,s})}{\frac{S}{K}\sum_{k=1}^{K}r^b_{i,k}} - 1,
\end{equation}

where $(r^b_{i,k})_{k\in\mathcal{K}}$ is the sequence of $r^b_{i,k}$ for all time slots and $\mathcal{K}=\{1,\cdots,K\}$.
Since playback 
continuity depends only on the base layer segments, it is possible that 
enhancement layers are fully delivered after the respective base layer is 
played. In other words, playback does not ``wait" for enhancement 
layers to arrive, while it does so for base layers. In such 
scenarios, the waiting time for base layers is the re-buffering 
time. From now on, we refer to events where the enhancement layers 
arrive late as a \textit{segment loss}, which is depicted in Figure 
\ref{fig:sl_late}. Enhancement layers that are delivered after the base layer 
playback are discarded and do not contribute towards the aggregate 
video quality. Based on this discussion, any scheme designed for optimal streamloading should take into account both re-buffering and enhancement layer segment loss. 

In addition to the constraints discussed above, there are other restrictions which further constrain the decision parameters for quality selection. For instance, users cannot request enhancement layers for segments that are already played back. Also, because of the base layer restrictions of streamloading, no base layer segments can be requested by users who have filled their buffer with base layer segments up to the buffer limit.

The joint resource allocation and quality selection for streamloading involves a complex interaction of decision variables which create several constraints on the system. In order to formulate the streamloading problem mathematically, we present OPTSL, which incorporates all the discussed constraints in an offline optimization framework. In order to do that, we define a set of non-negative auxiliary variables $x^b_{i,s,k}$ and $x^e_{i,s,k}$, which indicate the respective fraction of $r^b_{i,k}$ and $r^e_{i,k}$ used to download base and enhancement layers of segment 
$s$ in slot $k$, respectively. These auxiliary variables can be represented in vector form as $\mathbf{x}^b_{i,k}=(x^b_{i,s,k})_{s\in\mathcal S}$ and $\mathbf{x}^e_{i,k}=(x^e_{i,s,k})_{s\in\mathcal S}$.

We formulate the offline optimization problem OPTSL as follows:
\begin{align}\label{eq:offl_opt}
\nonumber
&\max_{\mathbf{r}^b_k,\mathbf{r}^e_k,\mathbf{q}_i,(\forall i\in \mathcal{N},k\in \mathcal{K})}  \sum_{i \in \mathcal{N}}\left(m_i^S(q_i)- \eta\text{v}^S_i(q_i)\right) \\\nonumber
&\text{subject to }\\	
&q_{i,s} \in \mathcal{Q}_{i,s},~  \forall s\in \mathcal{S},i \in \mathcal{N}\\
&x^{b}_{i,k,s}\geq 0,~x^{e}_{i,k,s}\geq 0,~  \forall s\in \mathcal{S},i \in \mathcal{N}\\
&x^{b}_{i,k,s}=0,~x^{e}_{i,k,s}=0 ~\forall k,s \label{const:causal} \\&~\mbox{s.t. }(k-\gamma_{i,k})\tau_{slot}>s\tau_{seg},i 
\in \mathcal{N} \nonumber\\
&||\mathbf{x}^{b}_{i,k}||_1\leq1,~||\mathbf{x}^{e}_{i,k}||_1\leq1~\forall k\in\mathcal{K},i \in 
\mathcal{N}  \label{const:seg_base}\\
&c_k(\mathbf{r}_k) \leq 0,~\forall k\in\mathcal{K} \label{const:capacity}\\
&r^b_{i,k} + r^e_{i,k} \leq r_{i,k},~\forall k\in\mathcal{K},i \in 
\mathcal{N} \label{const:rate_sum}\\
&\sum_{k=1}^K\tau_{slot}x^b_{i,k,s}r^b_{i,k}\geq\tau_{seg}f_{i,s}(q^0_{i,s})~\forall s \in \mathcal{S},i\in\mathcal{N} \label{const:fin_base}\\
&\sum_{k=1}^K\tau_{slot}x^e_{i,k,s}r^e_{i,k}\geq\tau_{seg}(f_{i,s}(q_{i,s})-f_{i,s}(q^0_{i,s})), \nonumber\\
&\forall s\in\mathcal{S},i\in\mathcal{N} \label{const:fin_enh}\\
&\beta_{i}(\mathbf{q}_i,(r^b_i)_{1:K}) \leq \bar{\beta}, \forall k\in\mathcal{K}, s\in \mathcal{S}, i \in \mathcal{N} \label{const:rebuf}\\
&\tau_{seg}S^b_{i,k} - (k-\gamma_{i,k})\tau_{slot}\leq \tau_{lim} , \forall k\in\mathcal{K}, i \in \mathcal{N} \label{const:buf_lim}\\
&\tau_{seg}S^e_{i,k}  - (k-\gamma_{i,k})\tau_{slot} \geq 0 , \forall k\in\mathcal{K}, i \in \mathcal{N},  \label{const:seg_loss}\\\nonumber
\end{align}
where $||\cdot||_1$ represents the $\ell^1$ norm.

In OPTSL, we define 
$S^b_{i,k}$ and $S^e_{i,k}$ as the total number of base and 
enhancement layers, respectively, that are completely delivered to 
user $i$ by slot $k$. The value of these two is derived as follows:

\begin{eqnarray}
S^b_{i,k} = \sum^S_{s=1}\mathds{1}_{\mathcal{A}^b_{i,s,k}}\left(r^b_{i,1},\cdots ,r^b_{i,k}\right),~\forall k\in\mathcal{K}, i \in \mathcal{N} \\
S^e_{i,k} = \sum^S_{s=1}\mathds{1}_{\mathcal{A}^e_{i,s,k}}\left(r^e_{i,1},\cdots ,r^e_{i,k}\right),~\forall k\in\mathcal{K}, i \in \mathcal{N}
\end{eqnarray}
where $\mathcal{A}^b_{i,s,k}$ and $\mathcal{A}^e_{i,s,k}$ are the set of rates that if allocated to user $i$, will allow the user to fully download the base and enhancement layers of segment $s$ by slot $k$, respectively. Hence, we define $\mathcal{A}^b_{i,s}$ and $\mathcal{A}^e_{i,s}$ as follows:
\begin{align}
\mathcal{A}^b_{i,s,k}=&\left\{r^b_{i,1},\cdots ,r^b_{i,k}|\tau_{slot}\sum_{t=1}^kx^b_{i,t,s}r^b_{i,t}\geq\tau_{seg}f_{i,s}(q^0_{i,s})\right\}, \label{eq:base_num}\\
\mathcal{A}^e_{i,s,k}=&\left\{r^e_{i,1},\cdots ,r^e_{i,k}|\tau_{slot}\sum_{t=1}^kx^e_{i,t,s}r^e_{i,t}\geq\tau_{seg}(f_{i,s}(q_{i,s})\right.\nonumber\\
&-f_{i,s}(q^0_{i,s}))\bigg\}. \label{eq:enh_num}
\end{align}

Each time the allocated set of rates makes a full base or enhancement layer download possible, the total number of downloaded segments $S^b_{i,k}$ and $S^e_{i,k}$ are incremented. 

Finally, $\gamma_{i,k}$ is the cumulative 
number of time slots up to slot $k$ that user $i$ has spent 
re-buffering. It is calculated in a manner similar to (\ref{eq:base_num}) and (\ref{eq:enh_num}), as follows:

\begin{eqnarray}
\gamma_{i,k} = \sum^k_{t=1}\mathds{1}_{\{S^b_{i,t}\tau_{seg}<(t - \gamma_{i,t-1})\tau_{slot}\}}(S^b_{i,t}), \mbox{where} ~\gamma_{i,0}=0 \nonumber\\~\forall k\in\mathcal{K}, i \in \mathcal{N} \label{eq:inst_rebuf}
\end{eqnarray}

The constraint shown in (\ref{const:rate_sum}) ensures that the sum of base and enhancement layer rates does not exceed the total allocated rate in each slot. The causality constraint (\ref{const:causal}) ensures that segments are not downloaded if their respective playback time has passed. Constraints (\ref{const:fin_base}) and (\ref{const:fin_enh}) ensure that the total rate allocated for downloading a specific segment, regardless of the layer, should be at least equal to the size of the segment. The right hand side of (\ref{const:fin_enh}) contains a decision variable indicating how many enhancement layers should be downloaded for each segment. The constraint should hold for any feasible choice of the number of enhancement layers. In (\ref{const:rebuf}), the upper limit on the fraction of time the user is re-buffering is set to $\bar{\beta}$. In other words, all segments need to be downloaded within $1+\bar{\beta}$ times the duration of the video. Therefore, $\bar{\beta}$ can take values greater than -1. However, the feasibility of the problem depends on the 
choice of $\bar{\beta}$, especially for negative values.  

The buffer limitation on the base layer segments is captured in (\ref{const:buf_lim}). Using this constraint, we ensure that at every time slot, the number of base layer segments currently stored in the buffer does not exceed the limit. The amount of buffered video at any given time slot is calculated as the total downloaded video duration minus the amount of time spent on playback. In order to avoid the occurrence of enhancement layer loss due to late arrivals, we introduce the \textit{segment loss constraint} (\ref{const:seg_loss}), which makes sure that the downloading header for the enhancement layers never falls behind the playback header.

The above problem jointly optimizes rate allocation 
over $\mathbf{r}^b_k$ and $\mathbf{r}^e_k$, and quality selection over $\mathbf{q}_i$, with 
respect to all given constraints. The feasibility 
depends on the choice of $c_k(\mathbf{r}_k)$, $\bar{\beta}$, and the 
quality rate trade-off functions. However, the solution to 
this problem is complex and requires channel state information for all future time slots, as well as the quality values of all future segments, thus it is not possible to implement it in practice. In order to overcome this problem, a sub-optimal online algorithm satisfying the constraints of OPTSL will be designed. The result of this algorithm is upper bounded by the optimal solution of OPTSL. In Section \ref{sec:sol}, we propose this algorithm, called RAQUEL, that performs rate allocation and quality selection in an online fashion.

\section{Online Algorithm RAQUEL}\label{sec:sol}

In this section, we present a simple online algorithm called RAQUEL that 
performs rate allocation and quality selection for multi-user streamloading in wireless networks. RAQUEL is based on an approach that was adopted 
for DASH video delivery in \cite{joseph2013nova,joseph2013mean}. The authors 
formulate the problem of DASH-based video delivery to a set of users
in a wireless network in a similar setting. The formulation includes a subset of the 
constraints in OPTSL, namely the link capacity 
constraint (\ref{const:capacity}) and the re-buffering 
constraint (\ref{const:rebuf}). Based on this formulation, an online 
algorithm is developed called NOVA, which is then proved to achieve 
optimality in a stationary regime. The fundamental idea in NOVA is the 
concept of \textit{virtual buffer}, which estimates the Lagrange multipliers associated with the re-buffering constraint, and hence, determines the risk of re-buffering for each user. 

In NOVA, rate allocation is performed by the base station at the beginning of each time slot, and quality selection is performed by individual users whenever they request a new segment.
At every time slot, the rate vector that maximizes $ \sum_{i\in\mathcal{N}}{b_ir_i}$ is determined, subject to the link capacity constraint, where $b_i$ denotes the value of the virtual buffer for user $i$. Therefore, users with better channel states and higher virtual buffer (higher risk of re-buffering) are prioritized for rate allocation. After allocating rates, the virtual buffers for all users are incremented by an amount proportional to $\tau_{slot}$.

A user who finishes downloading a segment, requests the quality level of the next segment $s$ to be delivered by maximizing $q_{i,s} -\eta(q_{i,s}-m_{i,s})^2 - \frac{b_i}{1+\beta_{NOVA}} f_i(q_{i,s})$, where $m_{i,s}$ keeps track of the average video quality of the segments delivered to user $i$ up to segment $s$. This objective function implies that since higher quality downloads require longer delivery times, high quality segments should be requested only if the risk of re-buffering is low, otherwise the requested quality should be decreased. The user who has finished downloading a segment, updates its virtual buffer by decreasing it proportional to $\tau_{seg}$. By following this procedure and constantly updating the virtual buffer, the obtained video quality is maximized without violating a constraint called $\beta_{NOVA}$, on the fraction of time spent re-buffering.

We use NOVA to simulate the streaming and downloading service models in Section \ref{sec:sim}. In RAQUEL, we devise a similar 
strategy as will be explained in Section \ref{sec:rqsl}. Before going 
through the details of the algorithm itself, we first describe the 
procedure for delivering base and enhancement layer segments.

\begin{figure}[t]
\includegraphics[height=1.6in]{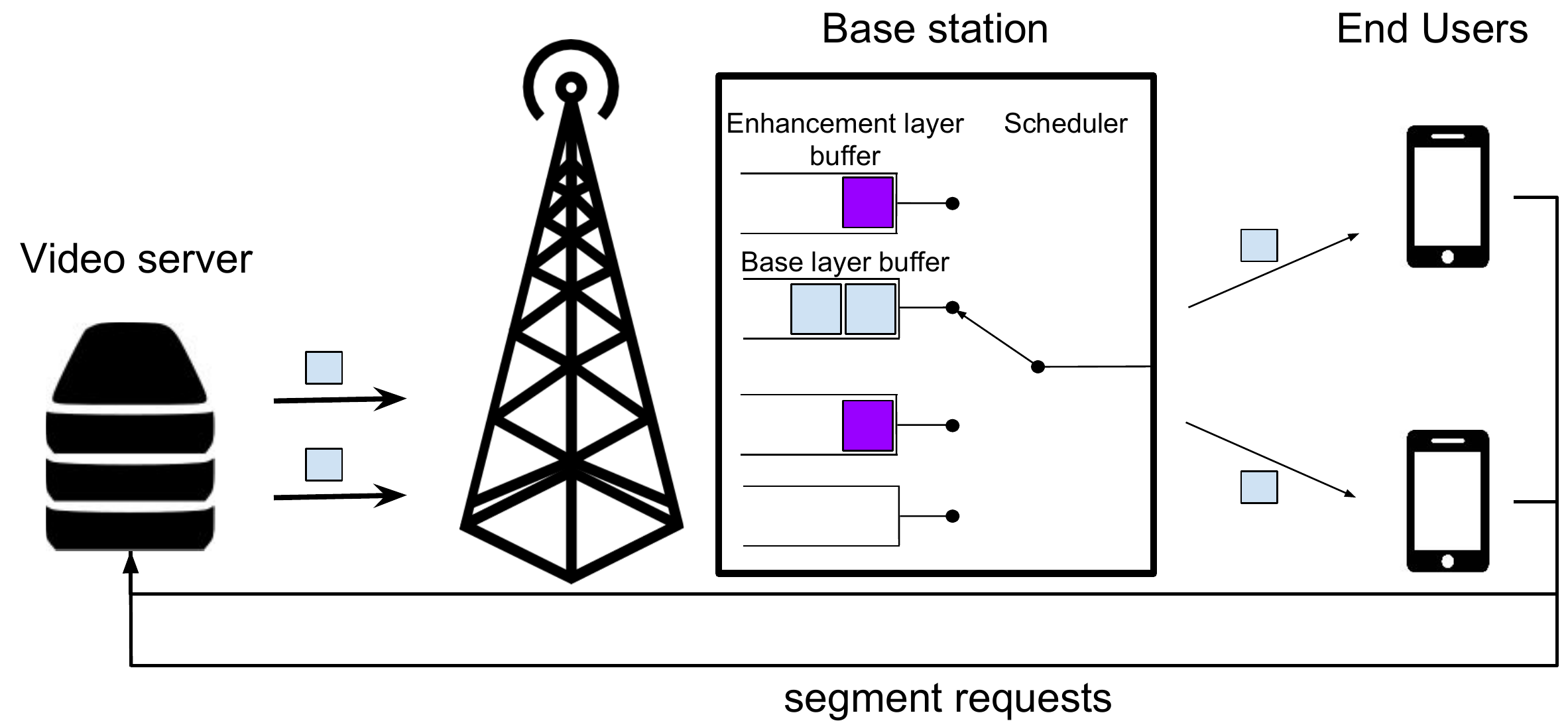}
\caption{The delivery procedure for streamloading with RAQUEL for two users. Each user is assigned two buffers at the base station for storing base and enhancement layer segments, respectively. Users request segments which the video server delivers to the base station and the base station schedules the segments for delivery.}
\label{fig:raquel}         
\end{figure}

\subsection{Delivery Procedure}\label{sec:proc}

In streamloading, base and enhancement layers are requested and delivered
separately. As described in Section \ref{sec:math_form}, segment requests 
in SVC-based video delivery are flexible in the sense that at any 
decision epoch, multiple layers of multiple segments can be requested by 
the user. Such flexibility allows for adaptive streaming schemes, like 
pre-fetching lower layers ahead and backfilling higher layers later 
\cite{hosseini2015not,andelin2012quality}. However, in designing RAQUEL, we 
assume that all requested enhancement layers of a particular segment are 
delivered together, and downloading additional enhancement layers for segments for which some
enhancement layers have already been downloaded in the past, is not possible. 

The segment delivery procedure that we propose in RAQUEL is illustrated in Figure \ref{fig:raquel}. We divide the joint rate allocation and quality selection into two separate tasks. The quality selection is done at the user end, where each user makes sequential requests on what segment to receive next. Upon receiving a request from a user, the video server sends the requested segment to the base station, which in turn schedules the users and delivers the segments. Here, we make the simplifying assumption that the end to end TCP connection between the server and the user can keep up with the segment delivery, hence, no congestion occurs at the link between server and base station.

In our proposed quality selection scheme, the users prioritize base layer over enhancement layer segments to reduce the likelihood of re-buffering. Users keep requesting base layer segments until they reach the buffer limit. Once the limit is reached, they request enhancement layers for the segments that have not been played back. The number of enhancement layers that each user requests for a particular segment is determined by the procedure explained in Section \ref{sec:adapt}. New enhancement layers are requested whenever the previously requested enhancement layers are fully delivered. New base layer segments are periodically requested as the playback continues and the buffered segments are consumed.

The rate allocation is done at the base station. As can be seen in Figure \ref{fig:raquel}, the base station assigns two buffers per user for the requested base and enhancement layer segments, respectively. The base station scheduler gives absolute priority to the base layer buffers and first tries to deliver all outstanding base layer segments to the users. Once no base layer segment is left at the base station, the scheduler starts allocating resources to deliver the buffered enhancement layer segments. The way the users are scheduled in each of these cases is explained in Section \ref{sec:alloc}.

By following the above procedure, segment quality selection and network resource allocation are independently and asynchronously performed by end users and the base station, respectively. Next, we explain each of these two steps in detail.

\subsection{RAQUEL}\label{sec:rqsl}

In this section, we explain the two tasks of RAQUEL, namely rate allocation at the base station (RA) and the quality selection at the user end (QUEL) in detail. Since the quality of each segment is a function of the number of layers requested for that segment, layer selection and quality selection are used interchangeably throughout the rest of the paper.

Similar to NOVA, we use two variables as the virtual buffer 
representation for the base and enhancement layers, which are dynamically
updated on a per slot basis and determine the allocated rate and 
selected quality for each user throughout the streamloading process. The 
virtual buffer for the base layer, $b^b$, is an indicator
of the risk of violating the re-buffering constraint (\ref{const:rebuf}).
A higher value for $b^b$ occurs whenever the occupancy of the base layer
buffer is low and hence the danger of re-buffering is 
high. Similarly, the virtual buffer for the enhancement layer $b^e$ indicates 
the risk of violating the segment loss constraint (\ref{const:seg_loss}). Due to 
the dynamic nature of the wireless channel, and also the varying buffer level at the user end, the virtual buffer values for both base and enhancement layers should be constantly updated. As the video plays back at the user end, the downloaded 
data in the buffer is consumed. Hence, as long as 
no new segment arrives at the user, the risk of 
draining the buffer constantly increases. Thus, at every time slot, the virtual 
buffer should increase proportional to the slot duration $b^{b}_{i} 
=b^{b}_{i} + \epsilon\tau_{slot}$ (same for $b^e_i$), where $\epsilon$ is a positive constant determining the rate of update.
However, if new segments are delivered to the user, the risk
of re-buffering (and segment loss) decreases proportional to the segment duration
and the corresponding virtual buffer is updated as $b^{b}_{i} 
=\max{\{b^{b}_{i} - \epsilon \tau_{seg},0}\}$ (same for $b^e_i$). The updated values are then used to perform RA and QUEL.

\subsubsection{Quality Selection QUEL}\label{sec:adapt}

As explained in Section \ref{sec:proc}, whenever user $i$ fully receives the enhancement layers it had requested for segment $s-1$, a decision has to be made about the quality level for the next
segment $s$. The request indicates how many enhancement layers user $i$ should download for segment $s$, according the following maximization:
\begin{align}\label{eq:qual_adapt}
\nonumber \textbf{QUEL}(b^e_i):&
\\l^*=\max_{l\in\{0,\cdots,L\}}&\bigg\{q^l_{i,s} - \eta(q^l_{i,s}-m_{i,s})^2 - \nonumber\\
&\left. \frac{b^e_i}
{1+\beta_{sl}}(f_i(q^l_{i,s})-f_i(q^0_{i,s})), i\in\mathcal{N}\right\}, 
\end{align}
where $q^{l}_{i,s}$ is the video quality user 
$i$ would see if it had $l$ enhancement layers in addition to the base layer, and $q^0_{i,s}$ is the minimum segment quality provided 
by the base layer for segment $s$. The average quality up to segment $s$ is denoted by $m_{i,s}$. It can be easily verified that the objective function is concave and solved by simply trying all possible levels $l$.

The right hand side of QUEL consists of three terms, where the 
second one accounts for the user sensitivity to variability in video quality, 
and thereby ensures that the requested quality is close to 
the average of the previously requested segments. The third term acts 
as a penalty on the requested quality to
avoid enhancement layer loss when the system is congested, i.e., $b^e_i$ is high. It can be observed 
that the larger the size of the segments become and the higher the risk
of starving the enhancement layer buffer gets, the more penalty is 
enforced on the requested segment quality. In addition to that, a key parameter in QUEL is $\beta_{sl}$ that aims at adjusting the 
sensitivity of the layer selection process to segment loss. For larger
$\beta_{sl}$ values, the system is less sensitive and as a result, tends to 
request more enhancement layers. This can result in a larger number of 
enhancement layer segment losses due to late arrival which leads to 
them being discarded. On the other hand, setting $\beta_{sl}$ to a low value 
increases the sensitivity to segment loss and results in a more 
conservative layer selection policy. Hence, changing the value of $\beta_{sl}$ has a two sided effect on the performance of QUEL.
In Section \ref{sec:sim} we analyze the impact of varying 
$\beta_{sl}$ on the overall video quality and enhancement layer segment 
loss, by showing that there exists an optimum value for $\beta_{sl}$ which trades off between aggressive layer selection and segment loss. A similar parameter, $\beta_{NOVA}$, is used in the NOVA algorithm. However, because of the fundamental differences
between streamloading and the other DASH based streaming schemes discussed
in Section \ref{sec:math_form}, $\beta_{NOVA}$ determines the sensitivity 
of quality selection on re-buffering. In fact, it can be easily verified 
that the two sided effect we observe with respect to $\beta_{sl}$ for 
streamloading does not hold with $\beta_{NOVA}$ for NOVA. Instead, increasing $\beta_{NOVA}$ will constantly increase the quality of 
requested segments while increasing the overall re-buffering time.

QUEL also takes into account the possibility of requesting no 
enhancement layers for a segment.
In such a scenario, user $i$ requests the minimum quality for segment $s$, and only the base layer will be delivered. Since 
no enhancement layer is requested for this segment, 
the enhancement layer 
download frontier shifts one segment ahead, followed by a request for 
segment $s+1$, according to 
(\ref{eq:qual_adapt}). This procedure repeats until the user 
requests one or more enhancement layers for a segment. 
To account for this, we update $b^{e}_{i} 
=\max{\{b^{b}_{i} - \epsilon \tau_{seg},0}\}$ as suggested before. In other words, we treat it as a \textit{complete download of zero 
enhancement layers}.

\subsubsection{Resource Allocation RA}\label{sec:alloc}
At the beginning of each time slot, the base station decides
how to allocate the available resources to each of the end 
users. As mentioned in Section \ref{sec:proc}, the scheduler keeps two buffers per user for base and enhancement layer segments, respectively. Resources are allocated with absolute prioritization of base layer segments. This means that if there are base layer segments left to transmit to the users at the base station, the base station schedules those first. This is done by solving $\text{RA}^b$ as shown below:

\begin{eqnarray}\label{eq:rate_alloc_base}
\textbf{RA}^\textbf{b}(b^b):\max_{\mathbf{r}} &\sum_{i\in\mathcal{N'}}{b^b_ir_i} 
\\\nonumber
s.t.& c(\mathbf{r})\leq0,i\in\mathcal{N'}
\end{eqnarray} 

where $\mathcal{N'}\subset\mathcal{N}$ is the set of all the 
users who have base layer data left at the base station. If there is no base layer segment left to transmit to the users, the scheduler allocates resources to deliver the queued enhancement layer segments by solving $\text{RA}^e$ as follows:

\begin{eqnarray}\label{eq:rate_alloc_enh}
\textbf{RA}^\textbf{e}(b^e):\max_{\mathbf{r}} &\sum_{i\in\mathcal{N}}{b^e_ir_i} 
\\\nonumber
s.t.& c(\mathbf{r})\leq0,i\in\mathcal{N}
\end{eqnarray} 

where $r_i$ is the allocated rate to user $i$. According to (\ref{eq:rate_alloc_base}) and 
(\ref{eq:rate_alloc_enh}), between users with the same 
achievable data rate, the one with larger virtual buffer has 
higher scheduling priority, and between users with equal virtual 
buffer, the one with higher achievable rate gets scheduled first.

Equations (\ref{eq:qual_adapt})-(\ref{eq:rate_alloc_enh}), capture how RAQUEL operates and Algorithm \ref{alg} shows a detailed description
including all the involved steps. RAQUEL is sub-optimal, but very simple to apply and does not require 
information about the future states of the channel. Furthermore, 
the allocation and layer selection steps can be independently and 
asynchronously performed at the base station and mobile station, 
respectively. 

\begin{algorithm}[h]
\caption{Streamloading Online Algorithm RAQUEL}
\label{alg}
\textbf{Initialization:} Let $\epsilon>0$ , and for each 
$i\in\mathcal{N}$, let $b_{i,0}^b\geq0$, $b_{i,0}^e\geq0$, and 
$0\leq m_{i,0}\leq q_{max}$.
\begin{algorithmic}
\FOR{all time slots k}
\STATE \textbf{ALLOCATE (RA):} 
\IF{Base layer segments present at base station}
\STATE $\textbf{RA}^{\textbf{b}}$ determines 
the optimal rate allocation vector $\mathbf{r}^b_k$.
\ELSE
\STATE $\textbf{RA}^{\textbf{e}}$ determines optimal rate allocation vector $\mathbf{r}^e_k$.
\ENDIF
\STATE Update virtual buffer as follows:
\begin{eqnarray}
b^b_{i,k+1} = b^b_{i,k} + \epsilon\tau_{slot} \label{eq:up_slot_b}\\
b^e_{i,k+1} = b^e_{i,k} + \epsilon\tau_{slot} \label{eq:up_slot_e}
\end{eqnarray}

\textbf{SELECT (QUEL):}
\IF{Base layer segments buffered up to the limit}
\FOR{$\forall i\in\mathcal{N}$}
\IF{user $i$ finished downloading enhancement layers for $s_i$}
\STATE Solve (\ref{eq:qual_adapt}) for user $i$ to obtain $l^*_{s_i+1}$.
\WHILE{ $l^*_{s_i+1}=0$}
\STATE
\begin{align}
m_{i,s_i+1} &=& &m_{i,s_i} + \epsilon(q^{0}_{i,s}-m_{i,s_i})^2,& \\
b^e_{i,k+1} &=& &\max{\{b^e_{i,k} - \epsilon \tau_{seg},0}\}\label{eq:up_seg_e},&\\
s_i &=& &s_i+1&
\end{align}
Solve (\ref{eq:qual_adapt}) for user $i$ to obtain $l^*_{s_i+1}$.
\ENDWHILE
\begin{align}
m_{i,s_i+1} &=& &m_{i,s_i} + (q^{l^*_{s_i+1}}_{i,s}-m_{i,s_i})^2,& \\
b^b_{i,k+1} &=& &\max{\{b^b_{i,k} - \epsilon \tau{seg},0}\}\label{eq:up_seg_b},&\\
s_i &=& &s_i+1&
\end{align}
\ENDIF
\ENDFOR
\ELSE
\FOR{$\forall i\in\mathcal{N}$}
\IF{user $i$ finished downloading current base layer}
\STATE Request next base layer 
\begin{equation}
b^b_{i,k+1}=\max{\{b^b_{i,k} - \epsilon \tau_{seg},0}\}
\end{equation}
\ENDIF
\ENDFOR
\ENDIF
\ENDFOR
\end{algorithmic}
\end{algorithm}

\section{Practical Implications}\label{sec:prac}
A goal for our streamloading algorithm is that its implementation be feasible on practical networks. Any algorithm for rate allocation needs to take into account the feasibility and practical limitations that exist on real base stations. Current base stations perform scheduling video streaming data at the MAC layer without considering the playback buffer state of each user. Such a cross-layer capability would increase the complexity of the base station design but offers substantial performance gains.
\begin{figure*}[t]
\centering
		\captionsetup[subfigure]{twoside}
		\begin{subfigure}[h]{0.29\textwidth}
             \includegraphics[height=1.6in]{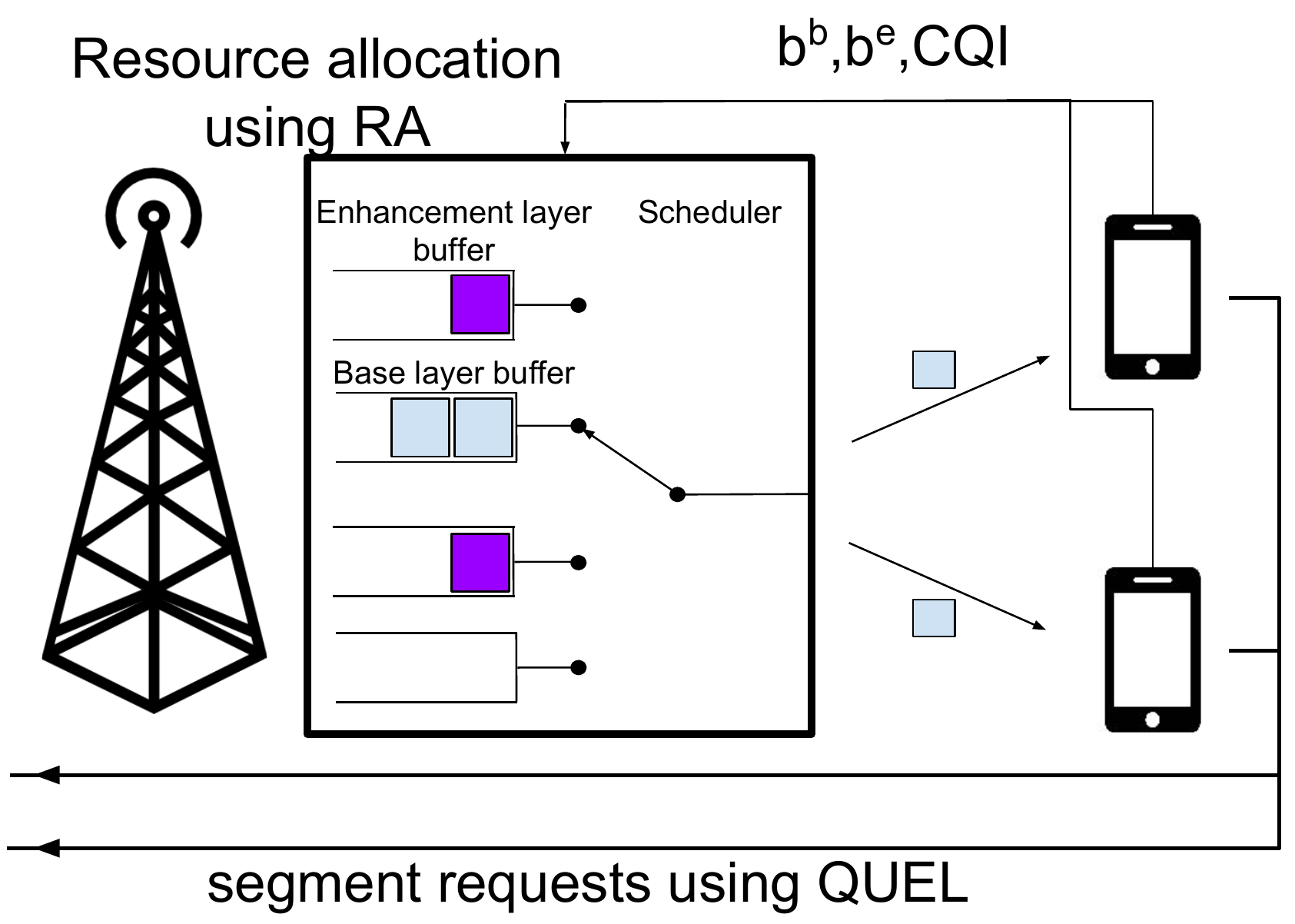}
             \caption{RA-based scheduling with two buffers per user at base station.}\label{fig:imp_raquel}         
           \end{subfigure}\qquad
           \begin{subfigure}[h]{0.29\textwidth}
             \includegraphics[height=1.6in]{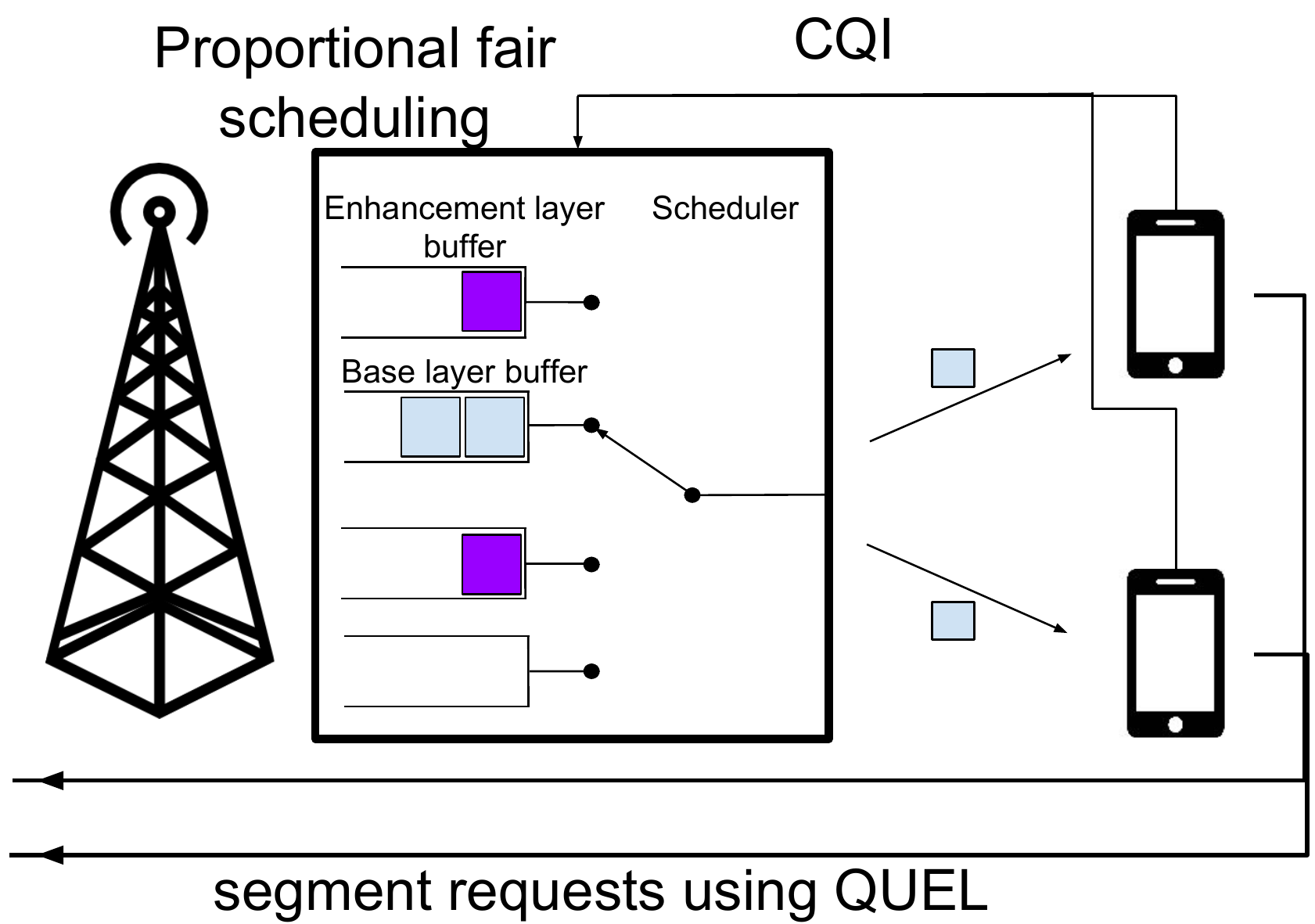}
             \caption{Proportional fair scheduling with two buffers per user at base station (PF1). }\label{fig:imp_pf}
           \end{subfigure}\qquad       
		\begin{subfigure}[h]{0.3\textwidth}
             \includegraphics[height=1.6in]{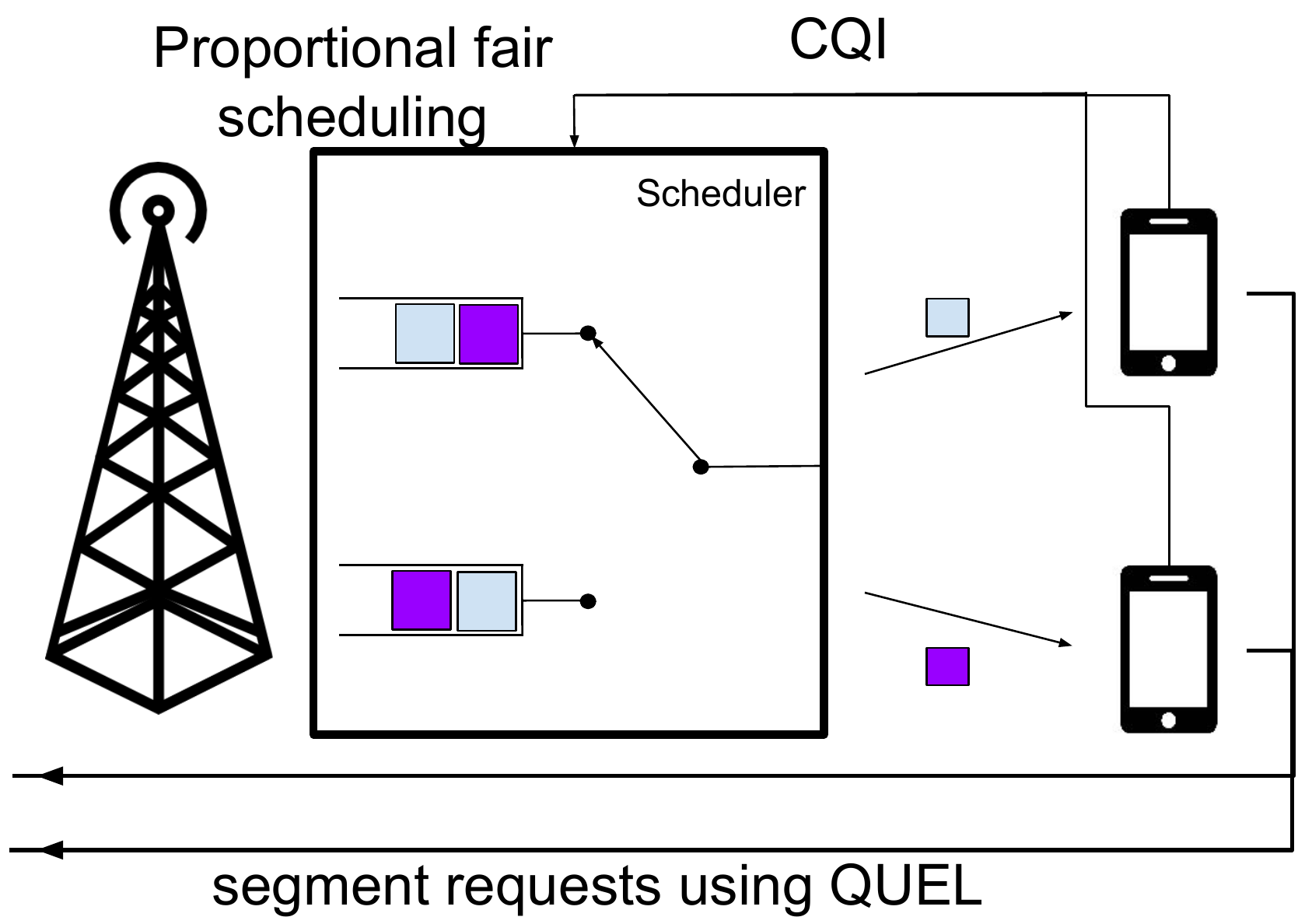}
             \caption{Proportional fair scheduling with single buffer per user at base station (PF1).}\label{fig:imp_simp_pf}         
           \end{subfigure}
\caption{Three different schemes for scheduling in terms of base station functionality. In all three cases, the quality request procedure is based on QUEL. From (a) to (c) complexity and signaling overhead decrease with the scheme presented in (c) not requiring any packet inspection at the base station.}
\label{fig:implementation}         
\end{figure*}

In this section, we evaluate practical implications emerging from implementing streamloading using RAQUEL in terms of signaling overhead, cross-layer requirements, and complexity. In order to compare RAQUEL with state-of-the-art scheduling mechanisms, we also present two schemes with decreased complexity and cross-layer functionality. 
All schemes follow the general procedure shown in Figure \ref{fig:raquel} and only the functionality of the scheduler changes as illustrated in Figure \ref{fig:implementation}. 

In Figure \ref{fig:imp_raquel}, the base station performs rate allocation using RA as explained in Section \ref{sec:alloc}, where two buffers are assigned to each user to store base and enhancement layer data, respectively. Hence, the base station needs to detect what layer the incoming packet belongs to, in order to properly implement RA. This can be done simply by either reading one bit from the application header of incoming packets or by using the standardized Differentiated Services (DiffServ)\cite{rfc24741998definition} field from the IP header. In addition to that, the user needs to send the updated values for the virtual buffer states $b^b$ and $b^e$ along with Channel Quality Indicator (CQI) reports back to the base station at the end of every time slot. Note that updates for $b^b$ and $b^e$, i.e., Equations (\ref{eq:up_slot_b}) and (\ref{eq:up_slot_e}) can be done independently at the base station, while Equations (\ref{eq:up_seg_e}) and (\ref{eq:up_seg_b}) require knowledge that a segment download has completed, which is easier to detect at the end users. The user can in turn send the updated values to the base station accordingly. Note that the latter only occurs at segment completion and is thus relatively infrequent. As a result, the overhead would be relatively low.

Figure \ref{fig:imp_pf} illustrates a similar system, with the difference that here the rate allocation is replaced by a simple proportional fair scheduler without the need of the virtual buffer values being fed back to the base station. Proportional Fairness (PF) is a simple standard scheduling mechanism for wireless networks in which in each time slot, the available rate is allocated to users in proportion to the average rate they have been allocated to date \cite{kelly1998rate}. Similar to the previous scenario, base layer data is prioritized over enhancement layer data during scheduling. We denote this scheduling scheme as PF1 for the remainder of the paper. The computational complexity of PF1 is similar to RA but the signaling overhead is reduced to only sending back CQI messages.

A third possible scheme is shown in Figure \ref{fig:imp_simp_pf}, where not only proportional fairness is used for scheduling, but also the base and enhancement layer buffers are replaced by a common buffer that is filled with data from different layers in the order in which they are requested. In this case, the benefits that result from prioritizing base layer over enhancement layer segments such as lower re-buffering time will be lost. In this scenario, no packet inspection is required by the base station. We denote this scheme as PF2 for the rest of the paper. The computational complexity and signaling overhead of PF1 is the same as PF2, but with reduced base station cross-layer functionality.

In Section \ref{sec:sim}, we evaluate the QoE achieved under deploying these three schemes and discuss the trade-offs that exist between introducing additional complexity and the enhancement in user's QoE.

\section{Simulation Results}\label{sec:sim}
In this section, we evaluate the performance of RAQUEL and compare it with the streaming and downloading 
service models via simulation. As discussed in Section \ref{sec:sol}, we implement NOVA to evaluate the performance of state-of-the-art streaming and downloading schemes. More precisely, since NOVA in its most general form does not have any buffer limit, it is best suited for the download service model. For the streaming scenario, when a user reaches the buffer limit, it stops 
requesting more segments until buffer space becomes available and NOVA resumes.

\subsection{Simulation Setting}\label{sec:sim_set}
The channel model under consideration follows capacity constraints in the 
form of 
$c_k(\mathbf{r}_k)=\sum_{i\in\mathcal{N}}\frac{r_{i,k}}
{\rho_{i,k}}-1$ in each time slot $k$, where $\rho_{i,k}$ is 
the maximum achievable rate for user $i$ in slot $k$. These peak 
rates are drawn from a rate distributions based on real HSDPA rate traces with correlation \cite{joseph2013nova}. 

The video to be delivered is a 20 minute long sequence from the 
open source Valkaama video which is divided into 1 second long 
segments. Each segment is encoded into 6 quality levels ranging from
100kbps to 1.5Mbps. For the DASH-based streaming and downloading 
scenarios, these different levels correspond to different quality 
representations, whereas for the SVC-based streamloading scenario, 
each of these levels corresponds to one additional enhancement 
layer. For example, consider a segment that is encoded into two representations 
of size 100kbps and 200kbps. We assume that the equivalent SVC representation of this segment consists of a base layer and one enhancement layer of equal
size. Hence, the quality obtained from downloading the 200kbps representation 
for DASH, is equal to the quality resulting from the base and one 
enhancement layer for SVC. The same holds true for additional 
enhancement layers. This example is a simplification of a real SVC video. Because of the overhead imposed by SVC, we add 
an extra $10\%$ to the size of each layer \cite{wiegand2007further}.

For each segment, we assume one base layer compressed at 100kbps and up to 
5 enhancement layers, the rates of which are shown in Table \ref{table:sim_param}.  In order for the quality-rate trade-off to capture the 
video quality that users perceive, we use a model for the Differential Mean Opinion 
Score (DMOS), see \cite{moorthy2010wireless}. 
In the absence of actual DMOS values, a proxy DMOS can be used to 
map each segment representation to the corresponding quality. Our proxy is based on the MSSSIM-Y metric for each 
segment according to a model presented in 
\cite{soundararajan2013video}. Our simulations are performed using the parameters in Table \ref{table:sim_param} unless stated otherwise.

\begin{table}[h] 
\centering
\begin{tabular}{|c|c|}
\hline
\textbf{Parameter} & \textbf{Value} \\
\hline
$\epsilon$ & 0.05\\
\hline
$\eta$ & 0.1 \\
\hline
$\tau_{slot}$ & 10ms \\
\hline
$\tau_{seg}$ & 1s \\
\hline
$\tau_{lim}$ & 50s \\ 
\hline
$\beta_{sl}$ & 0 \\
\hline
$\beta_{NOVA}$ & -0.2\\
\hline
video length & 20min\\
\hline
SVC overhead per enhancement layer& 10\% \\
\hline
number of enhancement layers & 5\\
\hline
base layer bit-rate & 100kbps\\
\hline
enhancement layer bit-rates & 
{100,100,300,300,600}kbps\\
\hline

\end{tabular}\caption{Simulation parameters}\label{table:sim_param}
\end{table}


\subsection{Improvements in Video Quality}
Our primary goal is to evaluate the performance of RAQUEL by comparing 
the resulting quality metrics of streamloading using RAQUEL with 
streaming and downloading. For this comparison, 
we evaluate our objective video quality metric, as well as the average re-buffering time. The video quality depends on the quality of each delivered segment minus a constant times the standard deviation of the video segment quality to account for users' negative response to variability \cite{yim2011evaluation}. Average re-buffering time is calculated as 
the cumulative amount of time that the video playback stops due to 
buffer starvation. 

\begin{figure}[h]
\includegraphics[height=0.6\linewidth]
{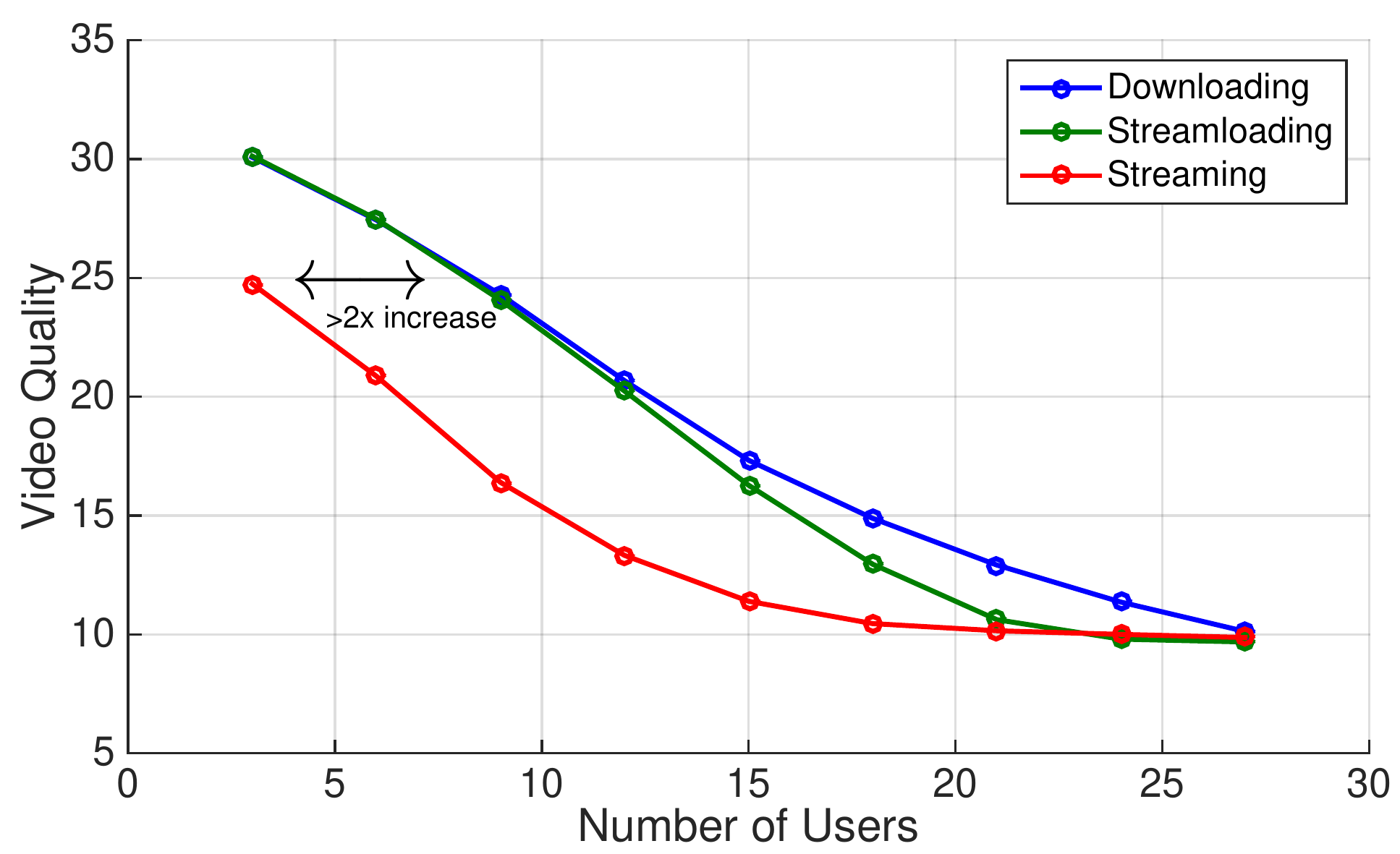}
\caption{Video quality comparison between streaming, downloading, and 
streamloading with a buffer limit equal to 50 seconds for 
streaming and streamloading.}\label{fig:qoe_comp_access}
\end{figure}

Figure \ref{fig:qoe_comp_access} shows the video 
quality obtained under each of the service models versus 
the number of users in the network. The buffer limit imposed on  
streaming and streamloading is set to 50 seconds. The figure 
shows that streamloading performs as well as unconstrained 
downloading for lightly loaded networks and at least as well as streaming for heavily loaded networks. It outperforms 
conventional streaming by a large margin, e.g., for a video quality of 25, the number of users that can be supported is doubled. This shows that 
streamloading provides video quality close to downloading, while 
still being legally classified as a streaming scheme.

\begin{figure}[h]     
\includegraphics[height=0.6\linewidth]
{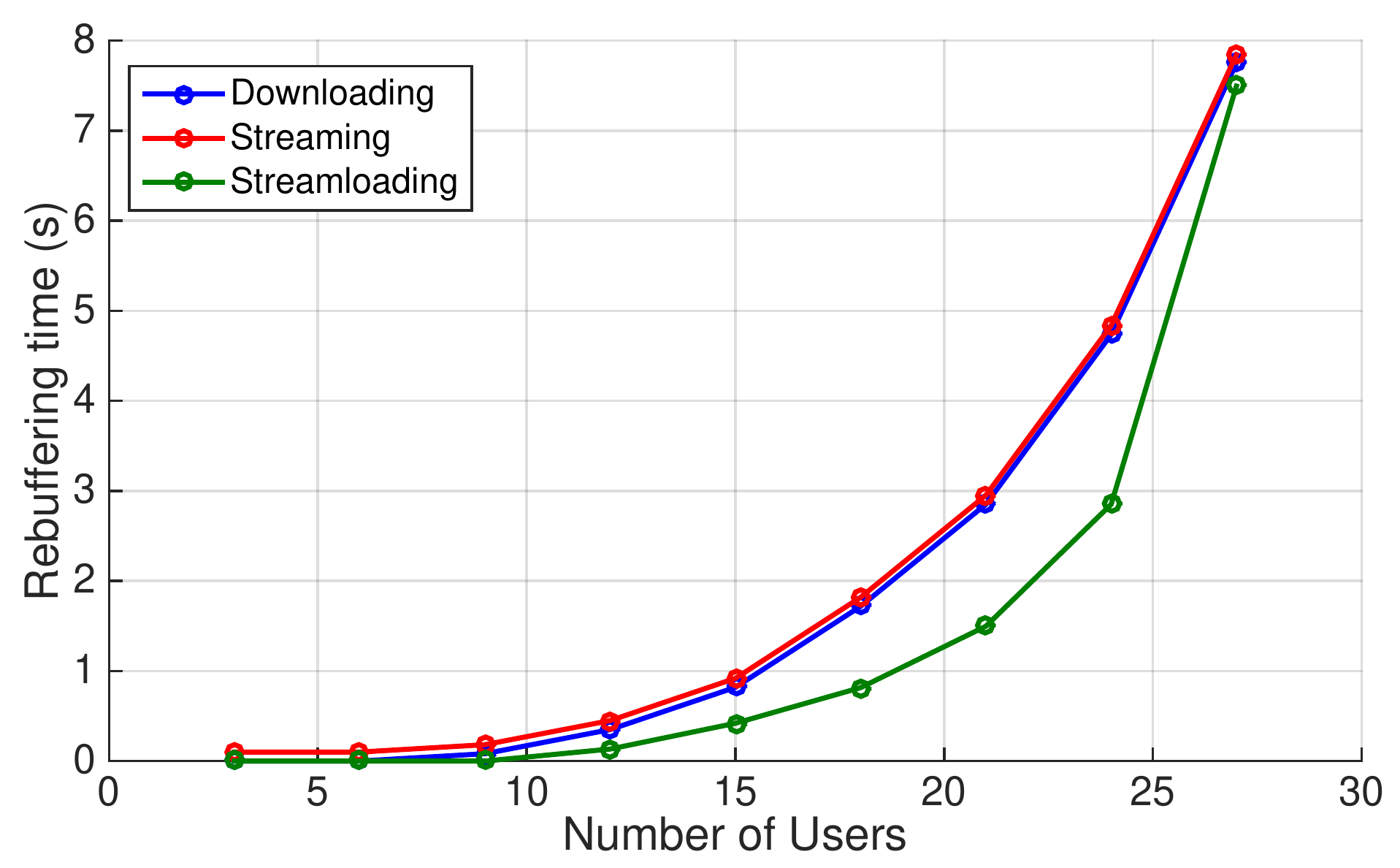}
\caption{Average re-buffering time for streaming, downloading, and 
streamloading with a buffer limit equal to 50 seconds for streaming 
and streamloading.}\label{fig:rebuf_comp_access}
\end{figure}

Figure \ref{fig:rebuf_comp_access} shows the average re-buffering 
time for each of the three service models. It can be seen that 
despite larger segment sizes due to encoding overhead, streamloading 
has shorter re-buffering times on average, as compared with streaming 
and downloading. Filling the base layer buffer prior to downloading 
enhancement layers is the reason for the lower re-buffering time. 
 
\begin{figure}[h]
\includegraphics[height=0.6\linewidth]
{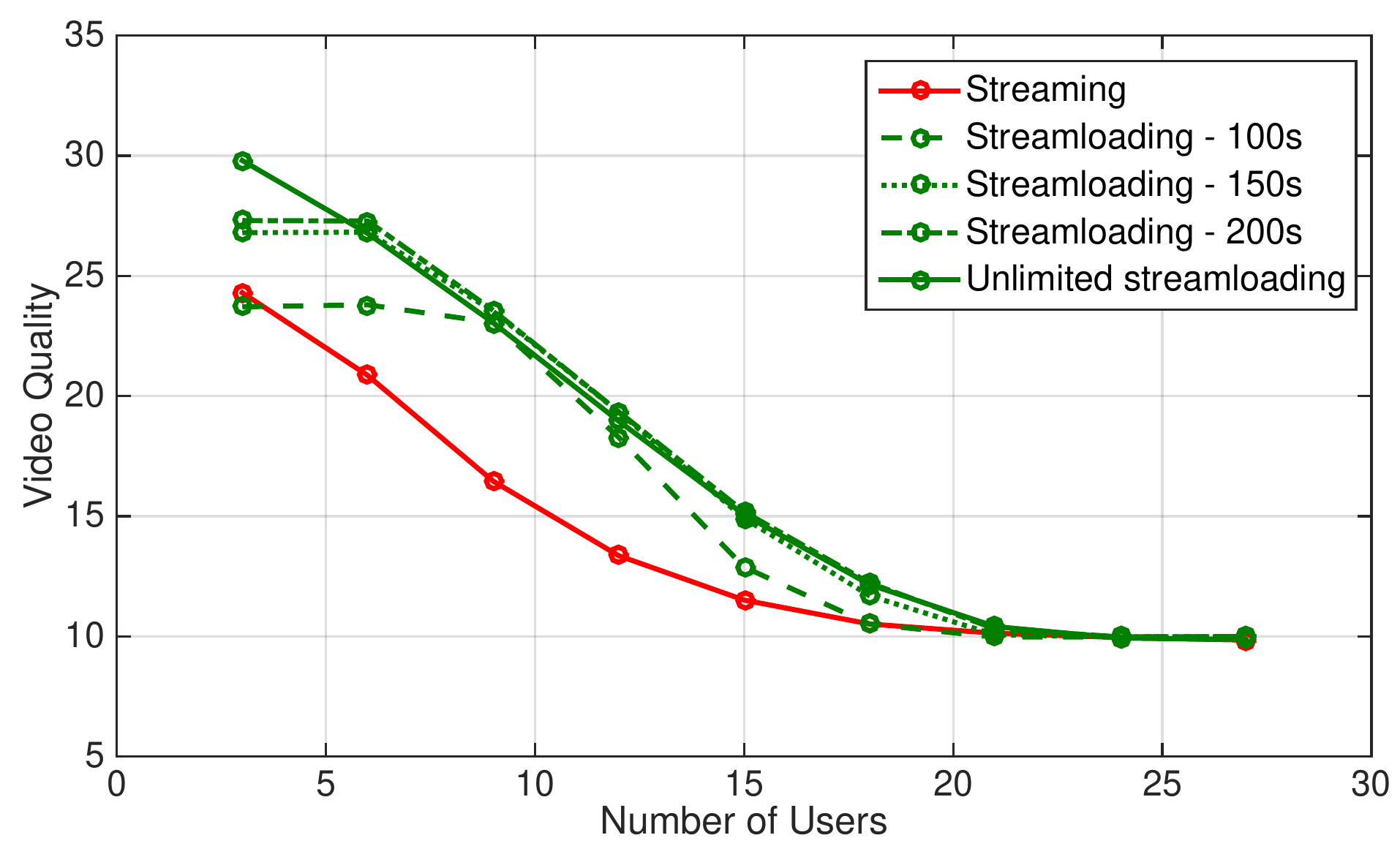}
\caption{Video quality comparison between streaming and streamloading with different limits for pre-fetching enhancement layer segments. The buffer limit for streaming as well as for base layers in streamloading is set to 50s.}\label{fig:buf_lim}
\end{figure}

A negative side effect of downloading too many segments ahead of playback is that if users stop watching the video before it ends (abandonment), the resources that are used to deliver the abandoned segments are wasted. The downloading and streamloading service models are prone to this wastage because of their pre-fetching functionality. However, streamloading has the advantage that the pre-fetched segments do not have the base layer, therefore, it causes less wastage of resources compared to downloading. This negative effect can be further mitigated if, similar to the base layer, a limit is set for pre-fetching enhancement layer segments. This limit should obviously be set to a larger value than the base layer buffer limit to gain the benefits  of streamloading. Figure \ref{fig:buf_lim} shows the video quality obtained from streamloading if the number of enhancement layers that can be pre-fetched is limited. It can be seen that even for an enhancement layer buffer limit of only 100s, the streamloading quality is higher than regular streaming. Furthermore, by setting this limit above 150s, streamloading performs almost as well as the case with unlimited pre-fetching. It should be noted that limiting pre-fetching to the values depicted in Figure \ref{fig:buf_lim} does not change the average re-buffering time. Hence, this limit can be set according to the trade-off between avoiding resource wastage and increasing video quality.

\begin{figure}[h]
\includegraphics[height=0.6\linewidth]
{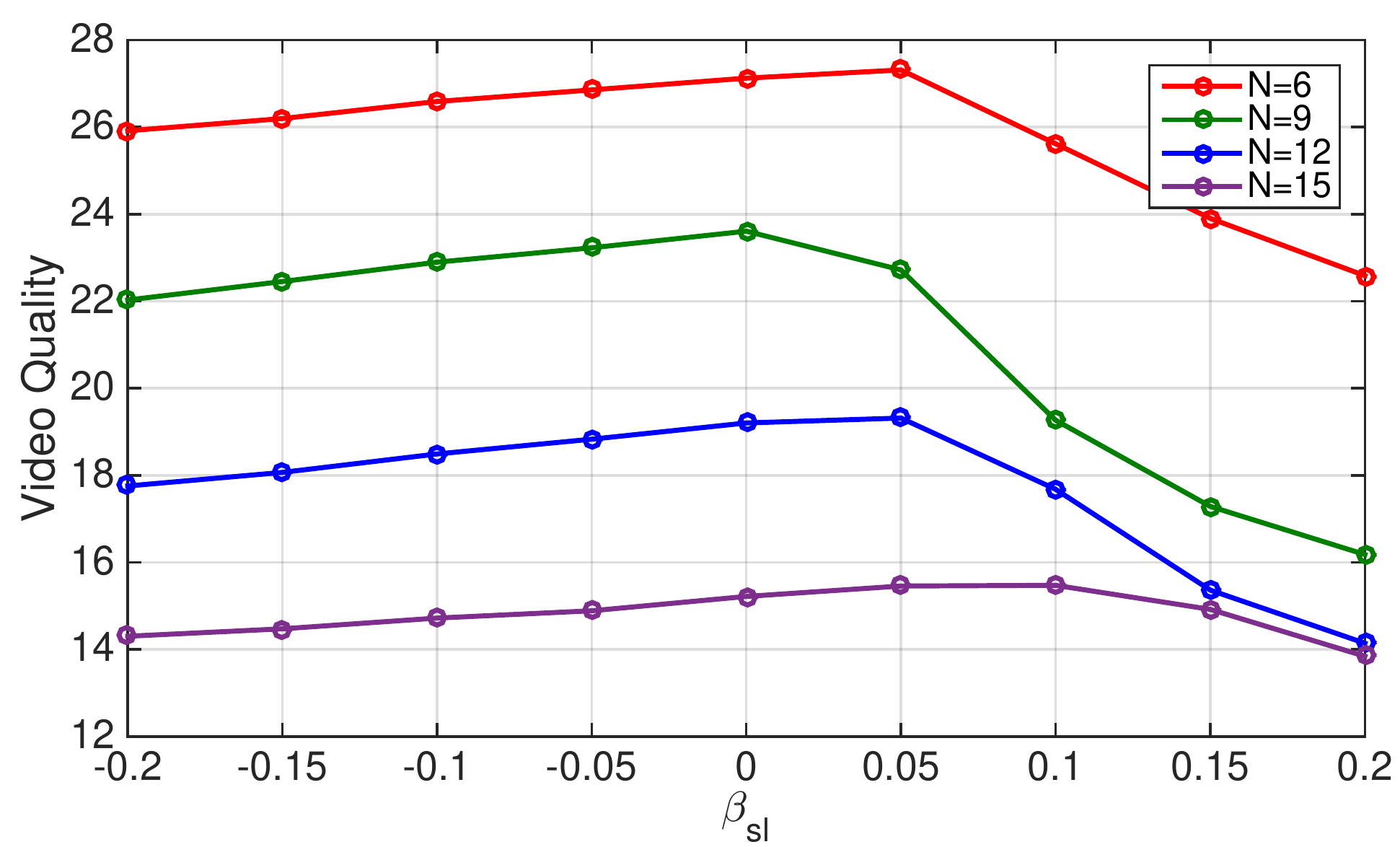}
\caption{Variations in video quality with respect to the segment loss sensitivity parameter $\beta_{sl}$ for different number of users. The trade-off between aggressive layer selection and segment loss, suggests an optimal value for $\beta_{sl}$ for different network loads.}\label{fig:qoe_beta}
\end{figure}

The two sided effect of $\beta_{sl}$ is shown in Figure \ref{fig:qoe_beta} under different network loads. As discussed in Section \ref{sec:adapt}, increasing $\beta_{sl}$ results in more aggressive layer selection which in turn increases the number of lost segments. According to this figure, up to a certain value for $\beta_{sl}$, the added aggressiveness results in higher video quality. However, increasing it further causes too many segment losses
which decrease the video quality and increase bandwidth wastage. The optimum value for $\beta_{sl}$ is roughly constant under various network loads. 

\subsection{RAQUEL vs. Baseline Algorithms}\label{sec:rqsl_comp}
Let us now evaluate the performance of RAQUEL by comparing it with two 
widely deployed algorithms, namely proportional fairness scheduling for resource allocation, and 
rate matching for quality selection. 
Rate Matching (RM), is a quality 
adaptation scheme in which the next selected segment is one whose 
bit rate is closest matching the average rate the 
user estimates it has seen to date \cite{akhshabi2012experimental}. For the rate allocation part, we investigate PF1 and PF2 as described in Section \ref{sec:prac}. 

\begin{figure}[h]     
\includegraphics[height=0.6\linewidth]
{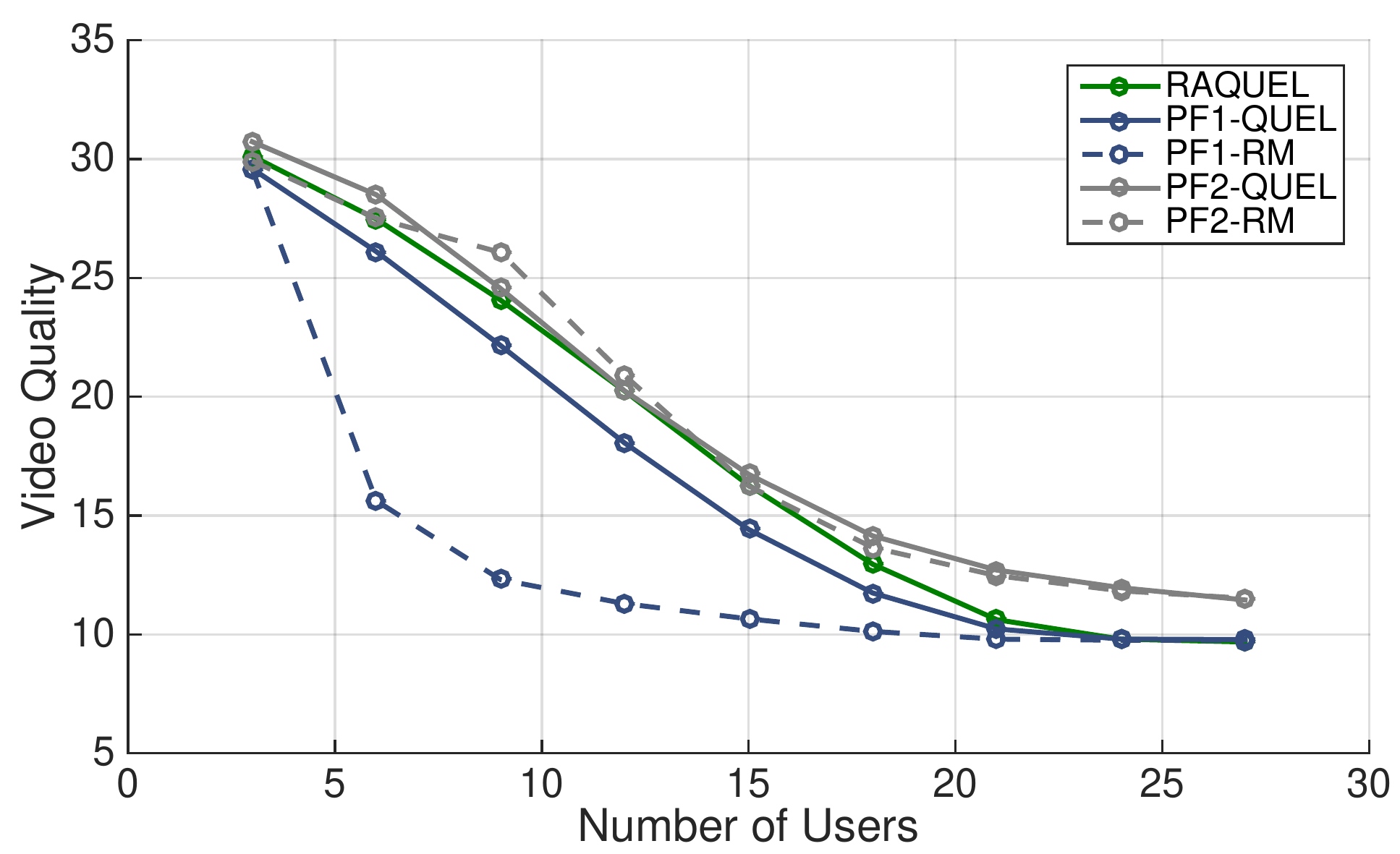}
\caption{Video quality comparison between RAQUEL, and the conventional proportional fairness and rate matching methods for streamloading. The base layer limit is fixed at 50 seconds.}\label{fig:qoe_alg}
\end{figure}

\begin{figure}[h]     
\includegraphics[height=0.6\linewidth]
{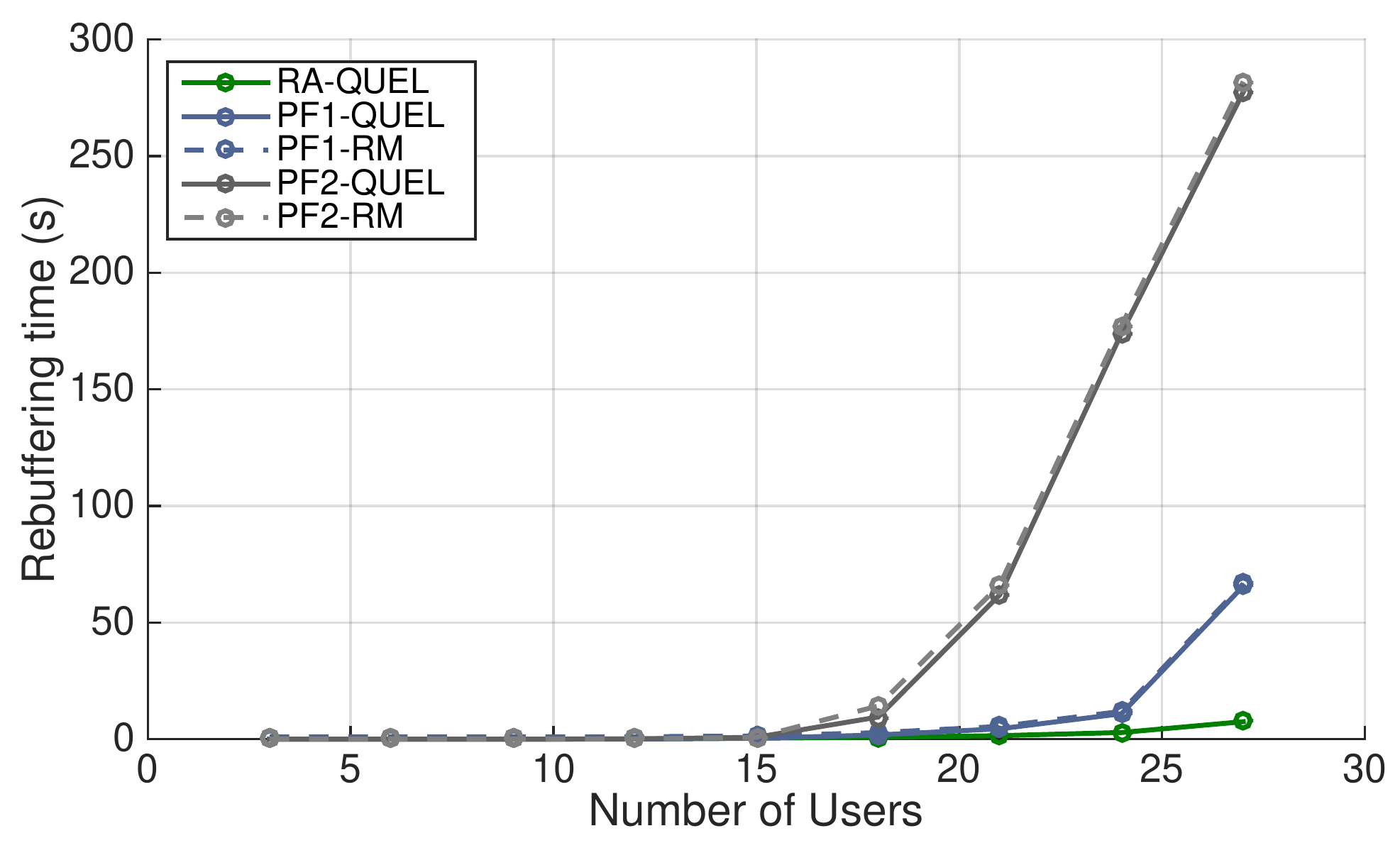}
\caption{Re-buffering comparison between RAQUEL and the conventional proportional fairness and rate matching methods for streamloading. The base layer limit is fixed at 50 seconds.}\label{fig:rebuf_alg}
\end{figure}

Figures \ref{fig:qoe_alg} and \ref{fig:rebuf_alg} demonstrate 
the performance of RAQUEL when compared to the cases where streamloading is done using the alternative schemes. For comparison, we 
adopted five different combinations for rate allocation and 
quality selection. In the first scheme, RAQUEL is 
applied to both tasks (RA-QUEL). The second method uses PF1 for rate allocation while QUEL is used 
for quality selection (PF1-QUEL). In the third approach, PF1 and rate matching replace RAQUEL in both tasks (PF1-RM). The fourth and fifth scheme are similar to the second and third where PF1 is replaced by PF2.  

As it can be seen from Figure \ref{fig:qoe_alg}, RAQUEL results in higher video quality than the two schemes that are based on PF1. However, the PF2-based algorithms perform slightly better than RAQUEL in terms of video quality. The reason for this is that since each user has a single buffer at the base station, scheduling is not done based on base layer prioritization. Hence, enhancement layer segments can be opportunistically downloaded and aggressively pre-fetched causing higher segment quality. However, not giving priority to base layer segments causes delay in their delivery  and consequently, results in re-buffering. Figure \ref{fig:rebuf_alg} shows that RAQUEL greatly outperforms all other schemes in terms of average re-buffering time. In fact, in PF2-based schemes, users are re-buffering almost 25\% of the entire streaming time. This shows that in PF1 and PF2, because scheduling is solely based on channel quality and the state of the buffer is not taken into account, re-buffering avoidance is not incorporated in the resource allocation, whereas in RAQUEL, the inclusion of the buffer level in the resource allocation mechanism through the virtual buffers reduces the re-buffering time.
However, PF1-QUEL may be an acceptable compromise on performance if lower complexity at the base station is an important consideration.

\section{Conclusions and Future 
Work}\label{sec:conc}
In this work, we have proposed an online procedure for 
asynchronous rate allocation and quality selection for  
streamloading. Streamloading is shown to provide low priced, high quality 
video to users watching copyright restricted content. This is done by pre-fetching enhancement layers ahead of real-time, and streaming base layer segments in real 
time. Our simulation results show that streamloading 
improves video quality over state-of-the-art streaming 
methods, while still satisfying the legal classification of a streaming service. 
Also, adding simple cross-layer functionality at the base station in order to distinguish between base and enhancement layer packets can enhance the QoE of the streamloading experience significantly.

The scope of this work can be further extended to 
different network types such as heterogeneous networks
consisting of high capacity femtocells or Wi-Fi 
hotspots, which can be leveraged for more efficient
pre-fetching of enhancement layers. 
Furthermore, the benefit of streamloading can be 
explored in a network with user dynamics, which 
includes users joining and leaving the network. The 
variability resulting from such traffic dynamics can be
exploited by speeding up the download of enhancement 
layers when the network is lightly loaded, in order to increase 
video quality when the network load is high. These
issues are subjects for future research.

\bibliographystyle{IEEEtran}
\bibliography{paper}
\end{document}